\documentclass[uft8]{article}

\pdfoutput=1 
\usepackage{url,hyperref,lineno,microtype}
\usepackage{multirow}
\usepackage{hyperref}
\usepackage[dvipsnames]{xcolor}
\usepackage{graphicx}
\usepackage{amsmath}
\usepackage{amsfonts}
\usepackage{amssymb}
\usepackage{natbib}
\usepackage{stmaryrd}
\usepackage{dsfont} % pour l'indicatrice
\usepackage{url}
\usepackage{upgreek} % lettre grec pas en diagonale
\usepackage{color, colortbl} % pour définir des nouvelles couleurs et pour colorer les tables

\definecolor{Gray}{gray}{0.9}

\newcommand{\printvalues}{topsep=\the\topsep; itemsep=\the\itemsep; parsep=\the\parsep; partopsep=\the\partopsep}

\DeclareMathOperator*{\argmin}{arg\,min}
\DeclareMathOperator*{\argmax}{arg\,max}

\newenvironment{tableth}{%
		\begin{table}[htbp]
			\centering
	}{
		\end{table}
		}

\graphicspath{{Images/}}
\RequirePackage[normalem]{ulem} 
\RequirePackage{color}\definecolor{RED}{rgb}{1,0,0}\definecolor{BLUE}{rgb}{0,0,1} 
\begin{document}
\onecolumn

\title{High heritability does not imply accurate prediction under the small additive effects hypothesis}

\author{Arthur Frouin\thanks{CNRGH, Institut Jacob, CEA - Universit\'e Paris-Saclay}, Claire Dandine-Roulland\footnotemark[1], Morgane Pierre-Jean\footnotemark[1],\\ Jean-François Deleuze\footnotemark[1], Christophe Ambroise\thanks{LaMME, Universit\'e Paris-Saclay, CNRS, Universit\'e d'\'Evry val d'Essonne} \thanks{These authors share last authorship} and Edith Le Floch\footnotemark[1] \footnotemark[3]} %This field will be automatically populated

\maketitle

\begin{abstract}

%%% Leave the Abstract empty if your article does not require one, please see the Summary Table for full details.
Genome-Wide Association Studies (GWAS) explain only a small fraction of heritability for most complex human phenotypes. Genomic heritability estimates the variance explained by  the  SNPs  on the whole  genome using mixed models and   accounts for the many small contributions of SNPs in the explanation of a phenotype.

This paper approaches heritability from a machine learning perspective, and examines  the close link between  mixed models and ridge regression.  Our contribution  is twofold. First, we propose estimating genomic heritability using a predictive approach via ridge regression and Generalized Cross Validation (GCV). We show that this  is consistent with classical mixed model  based estimation.
%This approach naturally establishes a statistical learning point of view on the concept of heritability.
Second,  we derive simple formulae that express  prediction accuracy as a function of the ratio $\frac{n}{p}$, where $n$ is the population size and $p$ the total number of SNPs. These formulae clearly show that a high heritability does not imply an accurate prediction when $p>n$.

Both the estimation of heritability via GCV and the prediction accuracy formulae are  validated using simulated data  and real data from UK Biobank.

\bigskip

\noindent
\textbf{Keyword -} heritability, prediction accuracy, ridge regression, mixed model, Generalized Cross Validation %

\end{abstract}

%%%%%%%%%%%%%%%%%%%%%%%%%%%%%%%%%%%%%%%%%%%%%%%
\section{Introduction}
%%%%%%%%%%%%%%%%%%%%%%%%%%%%%%%%%%%%%%%%%%%%%%%

The old nature versus nurture  debate is about whether a  complex human  trait is determined by a person's genes or by the environment. It is  a longstanding philosophical question that  has been  reinvestigated in the light of statistical genetics \citep{feldman1975heritability}. The concept of heritability was introduced by Sewall Wright \citep{wright1920relative,wright1921correlation} and Ronald Fisher \citep{fisher1919xv} in the context of pedigree data. It has  proved  highly useful in animal \citep{meuwissen2001prediction} and plant genetics \citep{xu2003estimating} for selection purposes because of its association with accurate prediction of  a  trait from genetic data. In the last decades, Genome-Wide Association Studies (GWAS) have become highly popular for identifying variants associated with  complex human  traits \citep{hirschhorn2005genome}. They  have recently  been used for heritability estimations \citep{yang_common_2010}. A shortcut is often made between the heritability of a trait and the prediction of this trait. However, heritable  complex human  traits are often caused by  a large number of genetic variants  that individually make  small  contributions to the trait variation. In this context, the relation between heritability and prediction accuracy may  not hold  \citep{de_vlaming_current_2015}.

\bigskip

The goal of this  paper  is to establish a clear  relation between prediction accuracy and heritability, especially when the number of genetic markers is much higher than the population size, which is typically the case in GWAS. Based on the linear model, statistical analyses of SNP data  address  very different and sometimes unrelated questions. The most  commonly performed analyses tend to  be association studies,    where multiple hypothesis testing  makes it possible to test the link between any SNP and  a phenotype of interest.  In genomic selection, markers are selected to predict a phenotype  with a view to selecting  an individual in a breeding population. Association studies and genomic selection may identify different sets of markers,  since even weak associations might be of interest for prediction purposes, while not  all strongly associated markers  are necessarily useful, because of  redundancy through linkage disequilibrium. Genomic heritability  allows quantifying  the  amount of genomic  information relative to a given phenotype via mixed model parameter estimation. The prediction of the phenotype using all genomic information via the mixed model is a closely related but different problem.

\bigskip

We approach the problem of heritability estimation from a machine learning  perspective. This is not a classical approach in genetics,  where  inferential statistics  is the  usual preferred tool . In this context, heritability is considered as a parameter   to be inferred from a small sample of the population.  The machine learning approach  places the emphasis on  prediction accuracy.
It  makes a clear distinction between performance on  training  samples and performance on testing samples,  whereas inferential statistics focuses on parameter estimation on a single dataset.% while not always assessing the precision of the estimator (bias and variance).

%-----------------------
\subsection{Classical Approach via Mixed Models}

Heritability is defined as the proportion of phenotypic variance due to genetic factors. A quantitative  definition of heritability requires a statistical model. The model commonly adopted is a simple three-term model without gene-environment interaction \citep{henderson_best_1975} :
$$
\textbf{y} = \textbf{g} + \textbf{f} + \textbf{e},
$$
where   $\textbf{y} \in \mathbb{R}^n$ is a quantitative phenotype vector describing $n$ individuals, $\textbf{f} \in  \mathbb{R}^n$ is a non-genetic covariate term, $\textbf{g} \in  \mathbb{R}^n$ is a genetic term and $\textbf{e} \in \mathbb{R}^n$ an environmental residual term. The term $\textbf{g}$ will depend on the diploid genotype matrix $\textbf{M} \in \mathcal{M}_{n,p} \left( \mathbb{R} \right)$ of the $p$ causal variants.
%whose $p$ variants are in linkage equilibrium (LE), as well as from a vector of genetic effects $\textbf{u} \in \mathbb{R}_p$.

\bigskip

There are two  definitions of heritability  in common use: first, there is $\text{H}^2$,  heritability in the broad sense, measuring  the overall contribution of the genome; and second, there is $\text{h}^2$,  heritability in the narrow sense  (also known as  additive heritability),  defined as the proportion of phenotypic variance explained by the additive effects of variants.

\bigskip

The quantification of narrow-sense heritability goes back to family studies by \citet{fisher1919xv}, who introduced the above model with the additional hypothesis that $\textbf{g}$ is the  sum of independent genetic terms, and with $\textbf{e}$ assumed to be normal. %Fisher defined the heritability in the context of family studies assuming independence  of genetic effects.
 This heritability in the narrow sense  is a function of the correlation between the phenotypes of relatives.
%When assuming additivity of the genetics effects within the Fisher model, we speak of heritability $h^2$ in a narrow sense, whereas heritability $H^2$ in the broad sense takes dominance into account.

%Also, it seems obvious, heritability is defined within a model assuming a form a mathematical link between phenotype an genotype (genetic architecture).

\bigskip

Although Fisher's original model makes use of pedigrees for parameter estimation,  some geneticists  have proposed using  the same model with genetic data from unrelated individuals \citep{yang_gcta:_2011}.

\subsubsection*{Polygenic model}

In this paper, we  focus on the version of the additive polygenic model with a Gaussian noise
%that is typically used to estimate heritability. Here $\psi \left( \textbf{g}, \textbf{f}, \textbf{e} \right) ±=   \textbf{g} + \textbf{f} + \textbf{e}$ and
where $\textbf{g} = \textbf{Z} u$, $\textbf{f} = \textbf{X} \beta$, with
$\textbf{Z} \in \mathcal{M}_{n,p} \left( \mathbb{R} \right)$ a standardized (by columns) version of $\textbf{M}$, $u \in \mathbb{R}^p$  a vector of genetic effects, $\textbf{X} \in \mathcal{M}_{n,r} \left( \mathbb{R} \right)$ a matrix of  covariates, $\beta \in \mathbb{R}^r$ a vector of covariate effects, $\mu$ an intercept and $\textbf{e} \sim \mathcal{N} \left( \textbf{0}_n, \sigma^2 \textbf{I}_{n} \right)$ a vector of environmental effects.

\bigskip

The model thus becomes
\begin{equation}
    \textbf{y} = \mu \boldsymbol{\mathds{1}}_n + \textbf{Z} u + \textbf{X} \beta + \textbf{e},
    \label{eq:modele_polygenique_base}
\end{equation}
where $\boldsymbol{\mathds{1}}_n \in \mathbb{R}^n$ a vector of ones.
%We  assume a standardized version of the phenotype for convenience, %but  show in the sequel  that centering is not necessary.

% In the additive polygenic model with Gaussian noise, the two definitions of heritability coincide $\text{H}^2 = h^2$.

\subsubsection*{Estimation of heritability from GWAS results}

To estimate heritability in a GWAS context, a first intuitive approach would be to estimate $u$ with a least squares regression to solve problem  (\ref{eq:modele_polygenique_base}). Unfortunately, this is complicated in practice for three reasons:  the causal variants are  not usually  available among genotyped variants; genotyped variants are in linkage disequilibrium (LD); and  the least squares estimate is only defined when $n>p$, which is not often the case in a GWAS \citep{yang_common_2010}.

\bigskip

One technique for  obtaining  a solvable problem is to use the classical GWAS approach to  determine  a subset of variants  significantly associated with the phenotype . The additive heritability can then be estimated  by summing their effects estimated by simple linear regressions. In practice this estimation  tends to  greatly underestimate $h^2$ \citep{manolio_finding_2009}.  It  only takes into account variants that have passed the significance threshold after correction for multiple comparisons (strong effects) and does not capture the variants that are weakly associated with the phenotype (weak effects).

\subsubsection*{Estimating heritability via the variance of the effects}
\citet{yang_common_2010} suggest that most of the missing heritability comes from variants with small effects.
 In order to be able to  estimate the information carried by weak effects  they assume a linear mixed model  where the vector of random genetic effects follows a normal homoscedastic distribution $u \sim \mathcal{N} \left( 0_p, \tau \textbf{I}_p \right)$. %$h^2$Using  Best Unbiased Linear Predictor (BLUP) \citep{henderson_best_1975} to estimate $\tau$ and $\sigma^2$,
They propose  estimating  the variance components $\tau$ and $\sigma^2$, and defining genomic heritability as $h^2_G = \frac{p \tau}{p \tau + \sigma^2}$.
%The BLUP is view from the random effets point of view : we are not looking to estimate genetics effects but rather their variances.
An example of an  algorithm for estimating variance components is the Average Information - Restricted Maximum Likelihood (AI-REML) algorithm, implemented in  software  such as Genome-wide Complex Trait Analysis (GCTA) \citep{yang_gcta:_2011} or \texttt{gaston} \citep{gaston}.

%-----------------------
\subsection{A statistical learning approach via ridge regression\label{Ridge}}

The linear model is used  in  statistical genetics for exploring and summarizing the relation between a phenotype and one  or more genetic variants, and  it is also used  in predictive medicine  and  genomic selection for prediction purposes. When used for prediction, the criterion for assessing  performance is the prediction accuracy.

\bigskip

Although least squares linear regression is the baseline method for quantitative phenotype  prediction, it has some limitations. As mentioned earlier,  the estimator is not defined when  the number of descriptive variables $p$ is greater  than the number of  individuals $n$. Even when $n>p$, the estimator  may  be highly variable when the descriptive variables are correlated,  which is clearly the case in genetics.

\bigskip

Ridge regression is a penalized version of  least squares that can  overcome these limitations \citep{hoerl1970ridge}. Ridge regression is strongly related to the  mixed model and is prediction-oriented.

\subsubsection{Ridge regression}

The ridge criterion builds on the least squares criterion,  adding an extra penalization term. The penalization term is  proportional to the $\ell_2$ norm of the parameter vector. The proportionality coefficient $\lambda$ is also  called the penalization parameter.
The penalty tends to shrink the coefficients of the least squares estimator,  but never cancels them out. The degree of shrinkage is controlled by $\lambda$: the higher the value of  $\lambda$, the greater the shrinkage:
\begin{eqnarray}
    \hat{u}_{R} & = &  \argmin_{u} \left\Vert \textbf{y} - \textbf{Z} u \right\Vert^2_2 + \lambda \left\Vert u \right\Vert_2^2, \\
    & = &   \left( \textbf{Z}^T \textbf{Z} + \lambda \textbf{I}_p \right)^{-1} \textbf{Z}^T \textbf{y}, \label{eq:estimateur_ridge_primal}\\
    &= & \textbf{Z}^T \left( \textbf{Z} \textbf{Z}^T + \lambda \textbf{I}_n \right)^{-1} \textbf{y}.
    \label{eq:estimateur_ridge_dual}
\end{eqnarray}

% The derivation of (\ref{Ridge_general}) allows to obtain the ridge estimator (\ref{Estimateur_ridge_general}).

% \begin{equation}
%     \hat{\text{u}}_R \left( \lambda \right) = \left( \textbf{Z}^T \textbf{Z} + \lambda \Id{p} \right)^{-1} \textbf{Z}^T \text{y} = \textbf{Z}^T \left( \textbf{Z} \textbf{Z}^T + \lambda \Id{n} \right)^{-1} \text{y} \label{Estimateur_ridge_general}
% \end{equation}

Ridge regression can be seen as a \emph{Bayesian Maximum a Posteriori} estimation of the linear regression parameters  considering  a Gaussian prior with hyperparameter $\lambda$.

\bigskip

The estimator  depends on a $\lambda$  that needs  to be chosen.  In a machine learning framework, a classical procedure  is to choose the  $\lambda$  that minimizes  the squared loss over new observations.
%Although there are many heuristics to choose $\lambda$, most of the time one relies on cross validation.

% To make one's choice, one can test a whole grid of $\lambda$ and look at which values make the best predictions.

% Despite the constraint of having to adjust an hyperparameter, ridge regression allows linear modeling by keeping all the variables in the model.

%The addition of a penalty on the size of the effects makes it possible to avoid their explosions when the variables are too correlated between them. More conveniently,

\bigskip

The practical effect of the penalty term is to add a constant to the diagonal of the covariance matrix,  which   makes the matrix non-singular, even in the  case where $p > n$.
When the descriptive variables are highly correlated,  this  improves the conditioning of the
$\textbf{Z}^T \textbf{Z}$ matrix,  while reducing the variance of the estimator.

\bigskip

The existence theorem states that there always exists a value of $\lambda>0$ such that the Mean Square Error (MSE) of the ridge regression estimator (variance plus the squared bias) is smaller than the MSE of the Maximum Likelihood estimator \citep{hoerl1970ridge}. This is  because  there is always an advantageous bias-variance compromise  that  reduces the variance  without greatly  increasing the bias.

\bigskip

%%% EDITH & CHRIS Dire un truc sur la prédiction.

% Ridge regression got multiple promising properties for modeling :
% unlike OLS, it can be done in the case of "high-dimension" ( $n < p$ ) even with highly correlated variable.
Ridge regression also allows us to simultaneously estimate all the additive effects of the genetic variants without discarding any,  which reflects the idea that  all the variants make a small contribution.

\subsubsection{Link between mixed model and ridge regression}

%\subsubsection{Our idea : estimating the vector of effects using ridge regression + estimation of $h^2_G$ using penalization parameters.}

This paper builds on the parallel between BLUPs (Best Linear Unbiased Predictions) derived from the mixed model and   ridge regression \citep{meuwissen2001prediction}. The use of ridge regression in quantitative genetics has already been discussed \citep{de_vlaming_current_2015, de_los_campos_prediction_2013} %(citation these sur capacité predictive ridge).
We  look at  a machine-learning oriented paradigm for estimating the ridge  penalty  parameter, which provides us with a direct link to heritability.
There  is an equivalence between  maximizing the  posterior $p \left( u \vert \textbf{y} \right)$ and  minimizing  a ridge criterion \citep{bishop_pattern_2006}. The penalty  hyperparameter  of the ridge criterion $\lambda$ is defined as the ratio of the variance parameters of the mixed model:

\begin{equation}
    \argmax_{u} p \left( u \vert \textbf{y} \right) = \argmin_{u} \left\Vert \textbf{y} - \textbf{Z} u \right\Vert^2_2 + \lambda \left\Vert u \right\Vert_2^2 \text{ with } \lambda = \frac{\sigma^2}{\tau}.
    \label{eq:equivalence_ridge_mm}
\end{equation}

The relation between $\lambda$ and $h_G^2$ \citep{de_vlaming_current_2015}
is thus:
\begin{equation}
    h_G^2 = \frac{p}{p + \lambda} ~ ; ~ \lambda = p \frac{1 - h_G^2}{h_G^2}.
    \label{eq:lambda_to_h2g}
\end{equation}
%,~ \hat{\lambda} = p \frac{1 - \hat{h}_G^2}{\hat{h}_G^2}
%\begin{equation}
%            h_G^2 = \frac{ p \tau}{p \tau + \sigma^2},~ \lambda = \frac{\sigma^2}{\tau}.
%    \label{Lien_h2v_lambda}
%\end{equation}

%Choosing $\lambda$ by cross validation is a standard procedure. Unfortunately cross validation imply high computational cost.
\

%is different from \citep{yang_common_2010} since now we focus on minimizing a cost function to estimate $\uv$.
%Ridge regression got multiple promising properties for modeling :
%unlike OLS, it can be done in the case of "high-dimension" ( $n < p$ ) even with highly correlated variable.
%It also allow us to simultaneously estimate all the additive effects of the genetic variants without discarding any, following the idea of a small contribution of all the variants.

% Using  the value of  $\lambda$  minimizing GCV, we also consider two alternative ways for estimating  the additive heritability from a prediction measure :

\subsubsection{Over-fitting}

% Edith Christophe
% Le message: la ridge utilise un jeu de donnée supplémentaire pour régler l'hyperparametre alors que le modèle mixte utilise une approche bayésienne empirique

% Voir Bishop:
%Over-fitting is a classical machine learning issue.

%Exhibiting too high performances on training data often impairs a good prediction of new test samples.  This phenomenon is known as   over-fitting.

%%%%% OVERFIT : complexite du modèle versus paramètres...

Interestingly, ridge regression and the mixed model can be seen as two similar ways to deal with the classical over-fitting issue in machine learning, which  is where a learner becomes  overspecialized in the dataset used for the estimation of its parameters and is  unable to generalize \citep{bishop_pattern_2006}.
When $n>p$, estimating the parameters of a  fixed-effect  linear model via  maximum likelihood estimation may lead to over-fitting,  when too many variables are considered.  A classical way of reducing  over-fitting  is regularization, and in order  to set the value of the regularization parameter  there are two commonly adopted approaches: first,  the Bayesian approach, and second,  the  use of additional  data.
%or equivalently resampling strategies.

\bigskip

Mixed  Model parameter  estimation via maximum likelihood can be seen as a type of self-regularizing approach (see Equation \ref{eq:equivalence_ridge_mm}).  Estimating  the variance components of the mixed model  may  be interpreted as a kind of empirical Bayes approach,  where the ratio of the variances is the regularization parameter that is usually  estimated  using a single  dataset. In contrast to this, in order  to properly estimate the ridge regression regularization hyperparameter that gives the best prediction, two datasets are required. If  a single dataset  were to be used, this would result in an insufficiently regularized (i.e., excessively  complex) model  offering  too high prediction performances on  the present dataset but unable to predict new samples well. This over-fitting phenomenon  is particularly evident when dimensionality is high.
%Two datasets are but required to avoid over-fitting in the estimation of the ridge regression regularization hyperparameter that gives the best prediction. Indeed, the use of a single dataset in machine learning would lead to the estimation of a too complex model which exhibits too high prediction performances on this dataset but  poorly predicts new samples. This over-fitting phenomenon is particularly visible in high dimension.

\bigskip

The fact that the complexity of the ridge model is controlled by its hyperparameter can be intuitively understood when considering extreme situations. When $\lambda$ tends  to infinity, the estimated effect vector (ie $\hat{u}_{R}$) tends to the null vector.  Conversely,  when $\lambda$ tends to zero, the model  approaches maximum complexity. One solution for choosing  the right complexity is therefore  to use both a training set to estimate the effect vector  for different values of the hyperparameter and a validation set to choose the hyperparameter value with the best prediction capacity on this independent sample.  An alternative solution, when data is  sparse, is to use  a cross-validation approach  to mimic a two-set situation. Finally, it should be noted that the estimation of  prediction performance on a validation dataset is still overoptimistic, and consequently  a third dataset,  known as  a test set, is required to assess the real performance of the model.

\subsubsection{Prediction accuracy in genetics}

In genomic selection and in genomic medicine, several authors  have been  interested in predicting complex traits that show a relatively high heritability using mixed model BLUPs \citep{speed2014multiblup}. The litterature  defined the prediction accuracy as the correlation between the trait and its prediction, which is unusual in machine learning where the expected loss is often preferred. Several approximations of this correlation  have been  proposed in the literature \citep{brard_is_2015}, either in a  low-dimensional  context (where the number of variants is lower than the number of individuals) or in a  high-dimensional  context.

\bigskip

\citet{daetwyler_accuracy_2008}
derived equations for predicting the accuracy of a genome-wide approach based on simple least-squares regressions for continuous and dichotomous traits. They
 consider one univariate linear regression  per  variant (with a fixed effect) and combine them afterwards, which is equivalent to a Polygenic Risk Score (PRS) \citep{pharoah2002polygenic,purcell2009o0donovan}.
%(PRS: polygenic risk scores).
 \citet{goddard_genomic_2009} extended this prediction to  Genomic BLUP (GBLUP), which used the concept of an  effective number of loci.
 %. His derivation builds on work by \cite{vi in which the variance of identical-by-descent sharing for full sibs was developed and provides a prediction for Me, which is Me 1⁄4 2NeL/ log(4NeL), where L is the genome length in Morgans (Goddard 2009).
\citet{rabier_accuracy_2016}  proposed   an alternative correlation formula conditionally on a  given training set. Their formula refines  the formula proposed by  \citet{daetwyler_accuracy_2008}. \citet{elsen_analytical_2017}  used  a Taylor development to derive the same formula in
small dimension.

\bigskip

Using intensive simulation studies, \citet{de_vlaming_current_2015} showed  a strong link between PRS and ridge regression in terms of prediction accuracy, when the population size is limited. However, with ridge regression, predictive accuracy improves substantially as  the sample size increases.

\bigskip

It is important to note a  difference in the prediction accuracy of GBLUP  when dealing with human populations as opposed to breeding  populations \citep{de_los_campos_prediction_2013}.
\citet{de_los_campos_prediction_2013} show that the squared correlation between GBLUP and the phenotype reaches the trait heritability, asymptotically when considering  unrelated human subjects.

\bigskip

\citet{zhao_cross-trait_2019} studied
cross trait prediction  in  high dimension. They derive generic formulae for in and out-of sample squared correlation. They link the marginal estimator to the ridge estimator and to GBLUP. Their results are very generic and generalize formulae proposed by \citet{daetwyler_accuracy_2008}.

\subsubsection{Outline of the paper}

While  some authors have proposed making use of  the equivalence between ridge regression and  the mixed model for setting the hyperparameter of ridge regression according to  the heritability estimated by the mixed model, we propose on the  contrary  to estimate the optimal ridge hyperparameter using a predictive approach via   Generalized Cross Validation. We derive approximations of the squared correlation and of the expected loss, both in high and low dimensions.

\bigskip

Using synthetic data and real data from  UK Biobank, we show that our results are consistent with classical mixed model based estimation and that our approximations are valid.

\bigskip

Finally,  with reference to the   ridge regression estimation of heritabilty, we discuss how  heritability is linked to  prediction accuracy in highly polygenic contexts.

%%%%%%%%%%%%%%%%%%%%%%%%%%%%%%%%%%%%%%%%%%%%%%%
\section{Materials and Methods}
%%%%%%%%%%%%%%%%%%%%%%%%%%%%%%%%%%%%%%%%%%%%%%%

%-----------------------
\subsection{Generalized Cross Validation for speeding up heritability estimation via  ridge regression }

\subsubsection{Generalized  Cross Validation }
A classical strategy for choosing the  ridge regression  hyperparameter  uses a  grid search and $k$ -fold  cross validation. Each grid value of the  hyperparameter  is evaluated by the cross validated error. This approach is  time-consuming  in high dimension,  since each grid value requires $k$ estimations.
In the machine learning context, we propose  using  Generalized Cross Validation (GCV) to speed up the estimation of the  hyperparameter  $\lambda$ and thus to estimate the additive heritability $h^2_G$ using the link described in Equation \ref{eq:lambda_to_h2g}.

\bigskip

The GCV error  in  Equation \ref{eq:def_GCV} \citep{golub_generalized_1978} is an approximation of the Leave-One-Out error (LOO) (see Supplementary Material). Unlike the classical LOO, GCV does not require $n$ ridge regression estimations (where $n$ is the number of observations) at each grid value, but involves a single  run. It thus provides a much faster and convenient alternative for choosing the  hyperparameter.  We have
% $$
% E_{LOO} &= \sum_{k=1}^n \left( \frac{ y_k - \hat{y}_{k} \left( \lambda \right) }{ 1 - h_{i,i} \left( \lambda \right) } \right)^2 \text{ and }
\begin{align}
    \text{err}^{GCV} &= \frac{\left\Vert \textbf{y} - \hat{\textbf{y}} \left( \lambda \right) \right\Vert^2_2}{ \left[ \frac{1}{n} \text{tr} \left( \textbf{I}_{n} - \textbf{H}_{\lambda} \right) \right]^2 },
\label{eq:def_GCV}
\end{align}

%where $\hat{\textbf{y}}\left( \lambda \right) = \textbf{Z} \hat{u}_{R} \left( \lambda \right) = \textbf{H}_{\lambda} \textbf{y}$ with \begin{align*}
 %   \textbf{H}_{\lambda} = & \textbf{Z} \left( \textbf{Z}^T \textbf{Z} + \lambda \textbf{I}_p \right)^{-1} \textbf{Z}^T \\
%= &\textbf{Z} \textbf{Z}^T \left( \textbf{Z} \textbf{Z}^T + \lambda \textbf{I}_n \right)^{-1},
%\end{align*}
where $\hat{\textbf{y}}\left( \lambda \right) = \textbf{Z} \hat{u}_{R} \left( \lambda \right) = \textbf{H}_{\lambda} \textbf{y}$ is the prediction of the training set phenotypes using the same training set for the estimation of $\hat{u}_{R}$ with
\begin{align*}
   \textbf{H}_{\lambda} = & \textbf{Z} \left( \textbf{Z}^T \textbf{Z} + \lambda \textbf{I}_p \right)^{-1} \textbf{Z}^T \\
= &\textbf{Z} \textbf{Z}^T \left( \textbf{Z} \textbf{Z}^T + \lambda \textbf{I}_n \right)^{-1}.
\end{align*}

\bigskip

A  Singular Value Decomposition (SVD) of the $\textbf{H}_{\lambda}$ can be used advantageously to speed up GCV computation  (see Supplementary Material).

\subsubsection{Empirical centering can lead to issues in the choice of penalization parameter in a  high-dimensional setting}

In high dimensional settings ($p>n$), the use of
GCV after empirical centering of the data can lead to a strong bias in the choice of $\lambda$ and thus in heritability estimation. Let us illustrate the problem with a simple  simulation. We simulate a phenotype from synthetic genotype data with a known heritability of $h^2=0.25$, $n=1000$ individuals, $p=10000$ variants and 100\%  causal variants. The simulation follows the  additive polygenic model without intercept or covariates, as described in Section \ref{sec:simulations}. Before applying GCV, genotypes are standardized in the most naive way : the genotype matrix $\textbf{M}$ is empirically centered and scaled column-wise, resulting  in the matrix $\textbf{Z}$.
Since we want to mimic an analysis on real data, let us assume that there is a potential intercept in our model (in practice  the empirical mean of our simulated phenotype is likely to be non-null):
%In most cases we would like to add an intercept in our model
\begin{equation}
    \textbf{y} = \mu \boldsymbol{\mathds{1}_n} + \textbf{Z} u + \textbf{e}.
    \label{eq:modele_polygenique_ac_intercept}
\end{equation}

\bigskip

GCV expects all the variables to be penalized,  but penalizing the intercept is not relevant. We therefore consider a natural two-step procedure: first  the model's intercept is estimated via the empirical mean of the phenotype $\hat{\mu} = \frac{1}{n} \sum_i y_i$, and, second,
 GCV is applied on the empirically centered phenotype $\textbf{y} = \textbf{y} - \hat{\mu} \boldsymbol{\mathds{1}_n}$.

\bigskip

\begin{figure}
    \centering
    \includegraphics[width=0.7\linewidth]{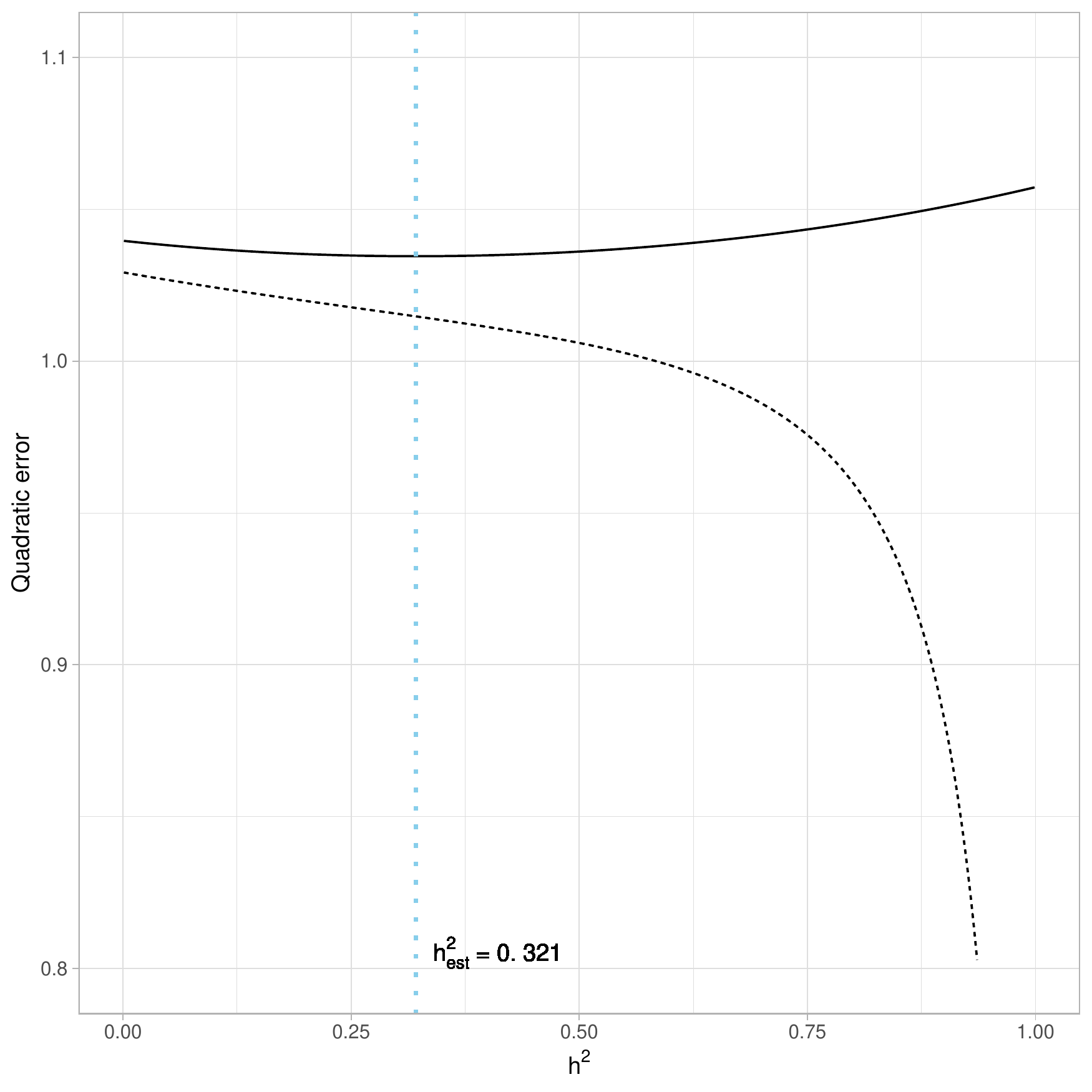}
    \caption{Example of biased estimation by GCV if $p>n$. We computed the GCV error curve with $n = 1000$ individuals, $p=10 000$ causal variants and simulated heritability $h^2_{sim} = 0.25$. We used a grid of $\lambda$ corresponding to the grid of heritability $\lbrace 0.01, 0.02, ..., 0.99 \rbrace$ using the link described in Equation \ref{eq:lambda_to_h2g} and computed the GCV error for those $\lambda$ after empirical standardization of the data (dotted line). The $\lambda$ that minimizes the GCV error corresponds to the heritability estimation. Here the GCV error tends to its minimum  as  $h^2$ tends to 1, and  heritability is thus largely over-estimated. The plain line corresponds to the GCV error obtained after correction of this bias by the projection approach (see Section \ref{projection}), which provides a  satisfactory  estimation of $h^2$.}
    \label{Problem_GCV}
\end{figure}

Figure \ref{Problem_GCV}  shows  the GCV error (dotted line).  Heritability is strongly overestimated.  The GCV error appears  to tend towards its minimum  as  $\lambda$  approaches  0 (i.e. when $h^2$ tends to 1).
%Figure \ref{Problem_GCV} displays the GCV error for two settings: $n>p$ ( $p=50$ and $n=1000$) and $n<p$ ( $p=10000$ and $n=1000$).  The vertical line shows the true value of the heritability. When $n>p$, GCV  reliably estimates the heritability, whereas there is a strong overestimation when $p>n$. In this case, the GCV error seems to tend towards its minimum when $\lambda$ goes to 0 (ie when $h^2$ tends to 1).

% An example based on simulation is shown in figure( see later sections for details on simulations ). Here we see that GCV highly overestimate the value of $h^2$ ( or underestimate the value of $\lambda$)  while random effects model provide satisfying results.

\bigskip

This  is a direct consequence of the empirical standardization of $\textbf{M}$ and of the phenotype.  By  centering the columns of $\textbf{M}$ with the empirical means of those columns, a  dependency is introduced,  and each line of the resulting standardized genotype matrix \textbf{Z} becomes a linear combination of all the others. The same  phenomenon of dependency  can be observed with the phenotype when using  empirical standardization.
Given the nature of the LOO in general (where each individual is considered successively as a validation set),  this kind of  standardization introduces a link between the validation set and the training set at each step: the ``validation set individual ''  can be written as a linear combination of the individuals in the training set.
%It thus seems obvious that this is particularly a problem in a high dimensional setting where overfitting occurs on the training set.
In high dimension, this  dependency  leads to $\text{err}^{LOO} \xrightarrow[\lambda \rightarrow 0] {}0$  (see Supplementary  Material ), due to over-fitting occurring in the training set.

\bigskip

From a GCV perspective, a related consequence of the empirical centering of the genotype data is that the matrix $\textbf{Z} \textbf{Z}^T$ has at least one null eigenvalue and an associated constant eigenvector in a high dimensional setting (see Supplementary Material).
%  Indeed these dependencies imply that the space of individuals can be described with only $n-1$ vectors (at most) in a high-dimensional setting.
This has a direct impact on GCV: using the singular value decomposition of the empirically standardized matrix $\textbf{Z} = \textbf{U} \textbf{D} \textbf{V}^T$ with $\textbf{U} \in \mathcal{O}_n \left( \mathbb{R} \right)$,  $\textbf{V} \in \mathcal{O}_p \left( \mathbb{R} \right)$ two orthogonal squared matrices spanning respectively the lines and columns spaces of $\textbf{Z}$ while $\textbf{D} \in \mathcal{M}_{n,p} \left( \mathbb{R} \right)$ is a rectangular matrix with singular values $\lbrace d_1, ..., d_n \rbrace$ on the diagonal. % , we showed in \ref{sec:limite_GCV} that
In a high dimensional context: $\text{err}^{GCV} ( \textbf{y} , \textbf{Z}, \lambda ) \xrightarrow[\lambda \rightarrow 0]{d^2_n = 0} ( \boldsymbol{\mathds{1}_n}^T \textbf{y} )^2 $. Performing the ``naive''  empirical centering of the phenotype results in
 \begin{align*}
     \text{err}^{GCV} ( \textbf{y} - \hat{\mu} \boldsymbol{\mathds{1}_n} , \textbf{Z}, \lambda ) &\xrightarrow[\lambda \rightarrow 0]{d^2_n = 0} ( \boldsymbol{\mathds{1}_n}^T \textbf{y} - \boldsymbol{\mathds{1}_n}^T \hat{\mu} \boldsymbol{\mathds{1}_n} )^2 = 0.
\end{align*}

\bigskip

The very same problem is observed for a more general model with covariates (see Supplementary Material).
%%Appendix section \ref{sec::pb_GCV_covariates}).

\subsubsection{A first solution using projection}
\label{projection}
%In addition to the dependency it creates, the empirical scaling before GCV may also be unsatisfying  because of the disjoint estimation of $\mu$ and $\textbf{u}$. We propose two solutions, one for each issue.

A better solution  for dealing  with the intercept (and a matrix of covariates $\textbf{X} \in \mathcal{M}_{n,r} \left( \mathbb{R} \right)$) in ridge regression is to use a projection matrix as a contrast and to work on the orthogonal of the space spanned by the intercept (and the covariates).
%Writing $\boldsymbol{\mathcal{X}} = \left[ \boldsymbol{\mathds{1}}_n, \textbf{X} \right]$
%\begin{align*}
%      \textbf{P}_{\boldsymbol{\mathcal{X}}} = \textbf{I}_n -  \boldsymbol{\mathcal{X}} \left( \boldsymbol{\mathcal{X}}^T \boldsymbol{\mathcal{X}} \right)^{-1} \boldsymbol{\mathcal{X}}^T
%\end{align*}

%The right multiplication of \ref{eq:modele_polygenique_base} by $\textbf{P}_{\boldsymbol{\mathcal{X}}}$ leads to a model without fixed effects but the estimation will account for it.% allowing a joint estimation of $\textbf{u}$ taking into account the intercept.

\bigskip

Contrast matrices are a commonly used approach in the field of mixed models for REstricted Maximum Likelihood computations ( REML ) \citep{patterson_recovery_1971}.  REML provides maximum likelihood estimation once fixed effects are taken into account. Contrast matrices are used to  ``remove''  fixed effects from the likelihood formula. If we are only interested in the estimation of the component of  variance, we do not even need to  make  this contrast matrix explicit : any semi-orthogonal matrix $\textbf{C} \in \mathcal{M}_{n-r-1,n}\left( \mathbb{R} \right)$ such that $\textbf{C} \textbf{C}^T = \textbf{I}_{n-r-1}$ and $\textbf{C} \times \left( \mu \boldsymbol{\mathds{1}_n} + \textbf{X} \beta \right) = 0_{n-r-1}$ provides a solution.
% The semi orthogonality of $\textbf{C}$ preserves the link ( \ref{Equivalence_ridge_mm} ) because the residuals keep their independence.
In a ridge regression context,  an explicit expression of $\hat{u}$ is needed for choosing the optimal complexity. An  explicit form for $\textbf{C}$ is therefore  necessary.

\bigskip

In the presence of covariates, a QR decomposition can be used to  obtain  an explicit form for $\textbf{C}$. %(see Supplementary Material)%Appendix section \ref{sec::QR_covariates}).
In the special case of an intercept without covariates, there is a  convenient choice of $\textbf{C}$. Since %we have shown (see Appendix \ref{sec::vecteur_propre_constant}) that
the eigenvector of $\textbf{Z} \textbf{Z}^T$ associated with the  final  null eigenvalue is constant, $\textbf{C} = \left[ \textbf{U}_1, ...,\textbf{U}_{n-1} \right]^T \in \mathcal{M}_{n-1,n} \left( \mathbb{R} \right)$ is a contrast matrix adapted for our problem. Additionally, by considering $\textbf{C} \textbf{Z}$ instead of $\textbf{Z}$, we have $\textbf{C} \textbf{Z} = \textbf{D}_{-n} \textbf{V}^T \rightarrow \textbf{C} \textbf{Z} \textbf{Z}^T \textbf{C}^T = \textbf{D}_{-n} \textbf{D}_{-n}^T$ with $\textbf{D}_{-n}$ the matrix $\textbf{D}$ deprived of row $n$. This choice of contrast matrix thus simplifies the GCV formula and allows extremely fast computation.

\subsubsection{A second solution using 2 data sets}

 Dependency between individuals can  be a problem when we use the same data for the standardization (including the estimation of potential covariate effects) and for the estimation of the genetic effects.  This can be overcome by partitioning  our data. Splitting our data into a standardization set and a training set, we will first use the standardization set to estimate the mean and the standard deviation of each variant, the intercept,  and the potential covariate effects. Those estimators will then be used to standardize the training set on which  GCV can then be applied .

% \begin{align*}
%      \forall j \in \llbracket 1,p \rrbracket ~ &\hat{m}_j = \frac{1}{n_{standard}} \sum_{i=1}^{n_{standard}} M^{standard}_{i,j} \\
%     & \hat{s}_j = \sqrt{ \frac{1}{n_{standard}} \sum_{i=1}^{n_{standard}} \left( M^{standard}_{i,j} - \hat{m}_j\right)^2 } \\
%     & \hat{\mu} = \frac{1}{n_{standard}} \sum_{i=1}^{n_{standard}} y_i \\
%     & \hat{\beta} = \left( \textbf{X}_{\textbf{standard}}^T \textbf{X}_{\textbf{standard}} \right)^{-1} \textbf{X}_{\textbf{standard}}^T \textbf{y}_{\textbf{standard}} \\
%     &\text{with } \textbf{X}_{\textbf{standard}} \text{ empirically scaled.}
% \end{align*}

% Then we compute $\text{GCV} \left( \textbf{y}_{\textbf{ridge}} - \hat{\mu} \boldsymbol{\mathds{1}}_{n_{ridge}} - e , \textbf{Z}_{\textbf{ridge}} \right)$

% QU'EST CE QUE L'ON FAIT ? EST-CE QUE TOUT PAR RAPPORT A BETA ESSAYE DE VENIR DE STANDART ? VA POSER DES PROBLEMES AVEC LES PC.

% \begin{align*}
%     \textbf{y}_{\textbf{ridge}},~\textbf{M}_{\textbf{ridge}} \rightarrow& \textbf{Z}_{\textbf{ridge}} = \left( \textbf{M}_{\textbf{ridge}} - \left[ \hat{m}_1 \boldsymbol{\mathds{1}}_{n_{ridge}}, ..., \hat{m}_p \boldsymbol{\mathds{1}}_{n_{ridge}} \right] \right) \times \textbf{diag} \left( \hat{s}_1, ..., \hat{s}_p \right) \\
%     & \textbf{y}_{\textbf{ridge}} = \textbf{y}_{\textbf{ridge}} - \hat{\mu} \boldsymbol{\mathds{1}}_{n_{ridge}}
% \end{align*}

%The use of different data sets for the standardization and GCV allow us to avoid dependency. We also do not have anymore any variable we wish not to shrink.

\bigskip

This method  has two main  drawbacks. The first  is that the estimation of the non-penalized effects is done independently of the estimation of the genetic effects, even though in practice we do not expect covariates to be highly correlated with variants. The other drawback is that it reduces the number of individuals for the heritability estimation (which is very sensitive to  the number of individuals). This approach therefore requires  a larger sample than  when using projection.

\subsection{Prediction versus Heritability in the context of small additive effects}

Ridge regression helps to highlight the link between heritability and
prediction accuracy. What  is the relation between the two concepts ?
Is prediction accuracy  an increasing function of heritability ? 

\bigskip

In a machine learning setting we have training and testing sets.
The index $_ {tr}$ refers to the training set, while  $_ {te}$ refers to the test set.

\bigskip

The classical bias-variance trade-off formulation considers the expectation of the loss over both the training set and test individual phenotype. It breaks down the prediction error into three terms commonly called variance, bias and irreducible
error.  In this paper we  do consider $\textbf{Z}_{tr}$ as fixed and the genotype of  a test individual as random, and somewhat abusively continue to employ the terms   variance, bias and irreducible error:
\begin{align*}
    \mathbb{E}_{\textbf{y}_{tr}, y_{te}, z_{te}} \left[ ( y_{te} - \hat{y}_{te} )^2 \right] &= \mathbb{E}_{z_{te}} \left[ \mathbb{E}_{\textbf{y}_{tr}, y_{te} \vert z_{te}} \left[ ( y_{te} - \hat{y}_{te} )^2 \right] \right] \\
                                                                                            &= \mathbb{E}_{z_{te}} \left[ \text{var}(y_{te} \vert z_{te}) + \text{var}(\hat{y}_{te} \vert z_{te} )\right] + \\
      &\ \mathbb{E}_{z_{te}} \left[ \left( \mathbb{E}_{\textbf{y}_{tr} \vert z_{te}} \left[ \hat{y}_{te} \right] - \mathbb{E}_{y_{te} \vert z_{te}} [ y_{te} ] \right)^2 \right].
\end{align*}

\bigskip

Assuming a training set genotype matrix $\textbf{Z} \in \mathcal{M}_{n,p} ( \mathbb{R} ) $ (without index $_{tr}$ to lighten notations) whose columns
have zero mean and unit variances, we denote
$\textbf{K}_{\lambda}= \left( \textbf{Z}^T \textbf{Z} +
  \lambda \textbf{I}_p \right)^{-1} \textbf{Z}^T$.
Assuming the independence of the variants
 $\mathbb{E}_{z_{te}} \left[ z_{te} \right] = 0_p$ and $\text{var}(z_{te}) = \textbf{I}_p$,
irreducible error, variance and bias become:
% \begin{align*}
%     \text{var}(y_{te} \vert z_{te}) &=  \sigma^2 \\
%     \text{var}(\hat{y}_{te} \vert z_{te} ) &=  \sigma^2 z_{te}^T \textbf{K}_{\lambda} \textbf{K}_{\lambda}^T z_{te} \\
%     \left( \mathbb{E}_{\textbf{y}_{tr} \vert z_{te}} \left[ \hat{y}_{te} \right] - \mathbb{E}_{y_{te} \vert z_{te}} [ y_{te} ] \right)^2 &= z_{te}^T \left( \textbf{K}_{\lambda} \textbf{Z} - \textbf{I}_{p} \right) u u^T \left( \textbf{K}_{\lambda} \textbf{Z} - \textbf{I}_{p} \right) z_{te}.
% \end{align*}.
\begin{align*}
    \mathbb{E}_{z_{te}} \left[ \text{var}(y_{te} \vert z_{te}) \right] &= \sigma^2 \\
    \mathbb{E}_{z_{te}} \left[ \text{var}(\hat{y}_{te} \vert z_{te} ) \right] &= \sigma^2 \text{tr} \left( \textbf{K}_{\lambda} \textbf{K}_{\lambda}^T \right) \\
    \mathbb{E}_{z_{te}} \left[ \left( \mathbb{E}_{\textbf{y}_{tr} \vert z_{te}} \left[ \hat{y}_{te} \right] - \mathbb{E}_{y_{te} \vert z_{te}} [ y_{te} ] \right)^2 \right] &= u^T \left( \textbf{K}_{\lambda} \textbf{Z} - \textbf{I}_{p} \right)^2 u.
% \\
%     &= u^T \left( \textbf{K}_{\lambda} \textbf{Z} \textbf{K}_{\lambda} \textbf{Z} - 2 \textbf{K}_{\lambda} \textbf{Z} + \textbf{I}_{p} \right) u.
\end{align*}
where $u$ is the vector of the ridge parameters.

\bigskip

Since  individuals are assumed to be unrelated, the  covariance matrix of the individuals is diagonal. The covariance matrix of the variants is also diagonal,  since   variants are assumed independent.
Assuming scaled data,  $\textbf{Z} \textbf{Z}^T$ and $\textbf{Z}^T \textbf{Z}$ are the empirical estimations of covariance matrices of respectively the individuals and the variants (up to  a $p$ or $n$ scaling factor).
Two  separate situations can be  distinguished according to the $n/p$ ratio.
In the high-dimensional case where $p>n$, the matrix $\textbf{Z} \textbf{Z}^T$  estimates well the individuals'  covariance matrix up to a  factor $p$. Where $n>p$, on  the other hand,  $\textbf{Z}^T \textbf{Z}$ estimates well the covariance matrix of variants up to a factor $n$.
Eventually,  $\textbf{Z} \textbf{Z}^T \simeq p \textbf{I}_n$ when $n < p$ and $\textbf{Z}^T \textbf{Z} \simeq n \textbf{I}_p$ when $n > p$.

\bigskip

Assuming further that
\begin{itemize}
    \item $\forall i \in \llbracket 1, n \rrbracket ~\text{var}(y_i) = 1$, we then have $\sigma^2 = 1 - h^2$ ,
    \item heritability is equally distributed among normalized variants i.e. $\forall j \in \llbracket 1, p \rrbracket ~\text{var}(u_j) = \frac{h^2}{p}$ (which is indeed the mixed model hypothesis),
    \item  $u^T u \simeq p \times \frac{h^2}{p}$ and $(\textbf{Z} u)^T (\textbf{Z} u) \simeq n h^2$,
\end{itemize}
the expected prediction error  can be stated more simply,  according to the $\frac{n}{p}$ ratio:
\begin{equation}
\mathbb{E}_{\textbf{y}_{tr}, y_{te}, z_{te}} \left[ ( y_{te} - \hat{y}_{te} )^2 \right]\simeq
\begin{cases}
  1 - \frac{n}{p} (h^2)^2,  \text{if} \  p\geq n\ \\
   (1 - h^2) \frac{1 + \frac{n}{p} h^2}{1 + h^2(\frac{n}{p} - 1)}, \  \text{otherwise}.
\end{cases}
\end{equation}

\bigskip

\begin{figure}
    \centering
    \includegraphics[width=0.7\linewidth]{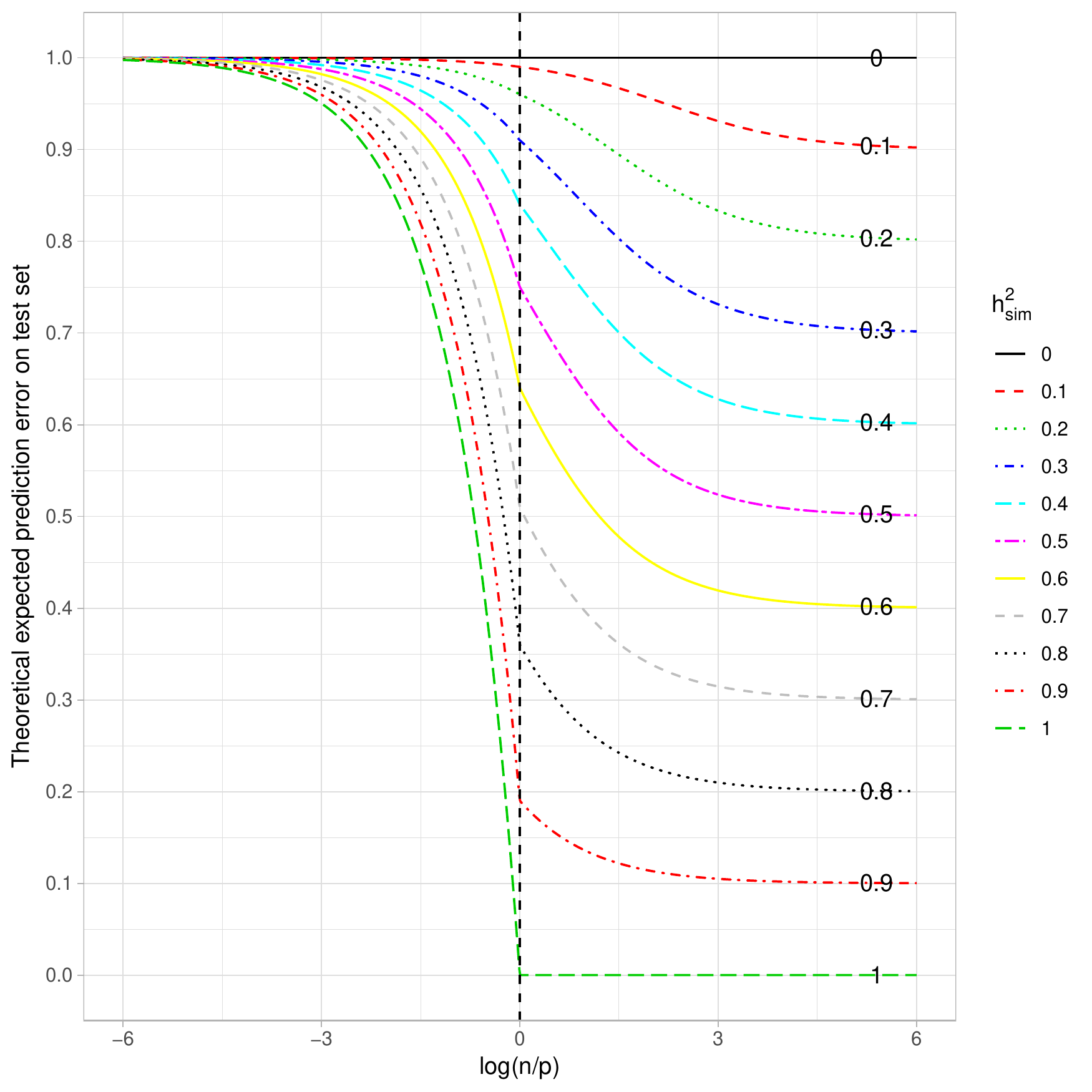}
    \caption{\label{fig:MSE-theo} Theoretical  Quadratic Error with respect to the log ratio of the number of individuals over the number of variants in the training set. Each curve corresponds to a given heritability (in the narrow sense).  Note that the total variance is assumed to be 1.}
\end{figure}

When considering the  theoretical quadratic error  with respect to  the log ratio of the number of individuals over the number of variants in the training set (Figure \ref{fig:MSE-theo}),  as expected we have  a decreasing function. This means that the larger the number of individuals in the training sample, the smaller the error. We also observe that the higher the heritability, the smaller the error. Both  of these things are intuitive, and as a consequence the  error tends towards the irreducible error when $n$ becomes much larger than $p$. What is more surprising is that the prediction error is close to the maximum, whatever the heritability, when $n$ is much smaller than $p$. Paradoxically, even with the highest possible  heritability, if the number of variants is too large  in relation  to the number of individuals,  no prediction is possible.

\bigskip

Similarly,  the prediction error can be computed on the training set instead of on  the test set. Using the same assumptions as before, the expected prediction error on the training set can be approximated by:
\begin{align*}
    \mathbb{E}_{\textbf{y}_{tr}} \left[ \frac{1}{n} ( \textbf{y}_{tr} - \hat{\textbf{y}}_{tr} )^T ( \textbf{y}_{tr} - \hat{\textbf{y}}_{tr} ) \right] \simeq \left\{ \begin{array}{l}
        ( 1 - h^2 )^2 \text{ if } p>n,  \\
        1 - 2 \frac{n}{n + \lambda} \left( \frac{p}{n} ( 1 - h^2 ) + h^2 \right) + \left( \frac{n}{n + \lambda} \right)^2 \left( \frac{p}{n} ( 1 - h^2 ) + h^2 \right) \text{ otherwise.}
    \end{array}\right.
\end{align*}

\bigskip

A graph similar to Figure \ref{fig:MSE-theo} for this expected error can be found in Supplementary Material. Interestingly,  when $p>n$, the error on the training set does not depend on the $n/p$ ratio. When $n$ becomes greater than $p$, it increases and tends towards the irreducible error $1 - h^2$ when $n \gg p$. As  shown  in Figure \ref{fig:MSE-theo}, the error on the test set is always higher than the irreducible error and thus higher than the error on the training set, which is a sign of  over-fitting . However, the difference between the error on the test set and the error  on the training set is a decreasing function of the $n/p$ ratio, which is linear when $p>n$ and tends towards zero when $n \gg p$.

\bigskip

Another popular way  of looking  at the predictive accuracy is to  consider  the squared correlation between $y_{te}$ and $\hat{y}_{te}$ \citep{daetwyler_impact_2010,goddard_genomic_2009}:
\begin{align*}
\text{corr}^2 ( y_{te}, \hat{y}_{te} )=\frac{\text{cov}^2 ( y_{te}, \hat{y}_{te} )}{\text{var} \left[ y_{te} \right]\text{var} \left[ \hat{y}_{te} \right]}.
\end{align*}

\bigskip

Although  correlation and  prediction error both provide information about the prediction accuracy,  correlation may have an interpretation that is intuitive, but it  does not take the scale of the prediction into account. From a predictive point of view, this is clearly a disadvantage.
% Although these two notions are very similar, the the correlation cannot detect if a scaling problem occurs in $\hat{y}_{te}$.
Considering $y_{te}$, $z_{te}$, and $y_{tr}$ to be random,  and
using the same assumptions  that were made in relation to  prediction error, the three terms of the squared correlation become:
\begin{align*}
    \text{cov}^2 ( y_{te}, \hat{y}_{te} ) &= (u^T \textbf{K}_{\lambda} \textbf{Z}_{tr} u)^2, \\
    \text{var} \left[ \hat{y}_{te} \right] &=\text{tr}( \textbf{K}_{\lambda}^T \textbf{K}_{\lambda} \times \sigma^2 \textbf{I}_n ) + ( \textbf{Z}_{tr} u )^T \textbf{K}_{\lambda}^T \textbf{K}_{\lambda} ( \textbf{Z}_{tr} u ),\\
    \text{var} \left[ y_{te} \right] &=1.
\end{align*}

\bigskip

Like in the case of  prediction error,  replacing $\textbf{Z} \textbf{Z}^T$ or $\textbf{Z}^T \textbf{Z}$ by their expectations, the squared correlation simplifies to:
\begin{align}
    \text{corr}^2 ( y_{te}, \hat{y}_{te}) \simeq \left\{ \begin{array}{l}
         \frac{n}{p}  (h^2)^2 \text{ if } n < p, \\
          \frac{(h^2)^2}{ \frac{p}{n} ( 1 - h^2 ) + h^2}  \text{ otherwise.}
    \end{array} \right.
\end{align}

\bigskip

\begin{figure}
    \centering
    \includegraphics[width=0.7\linewidth]{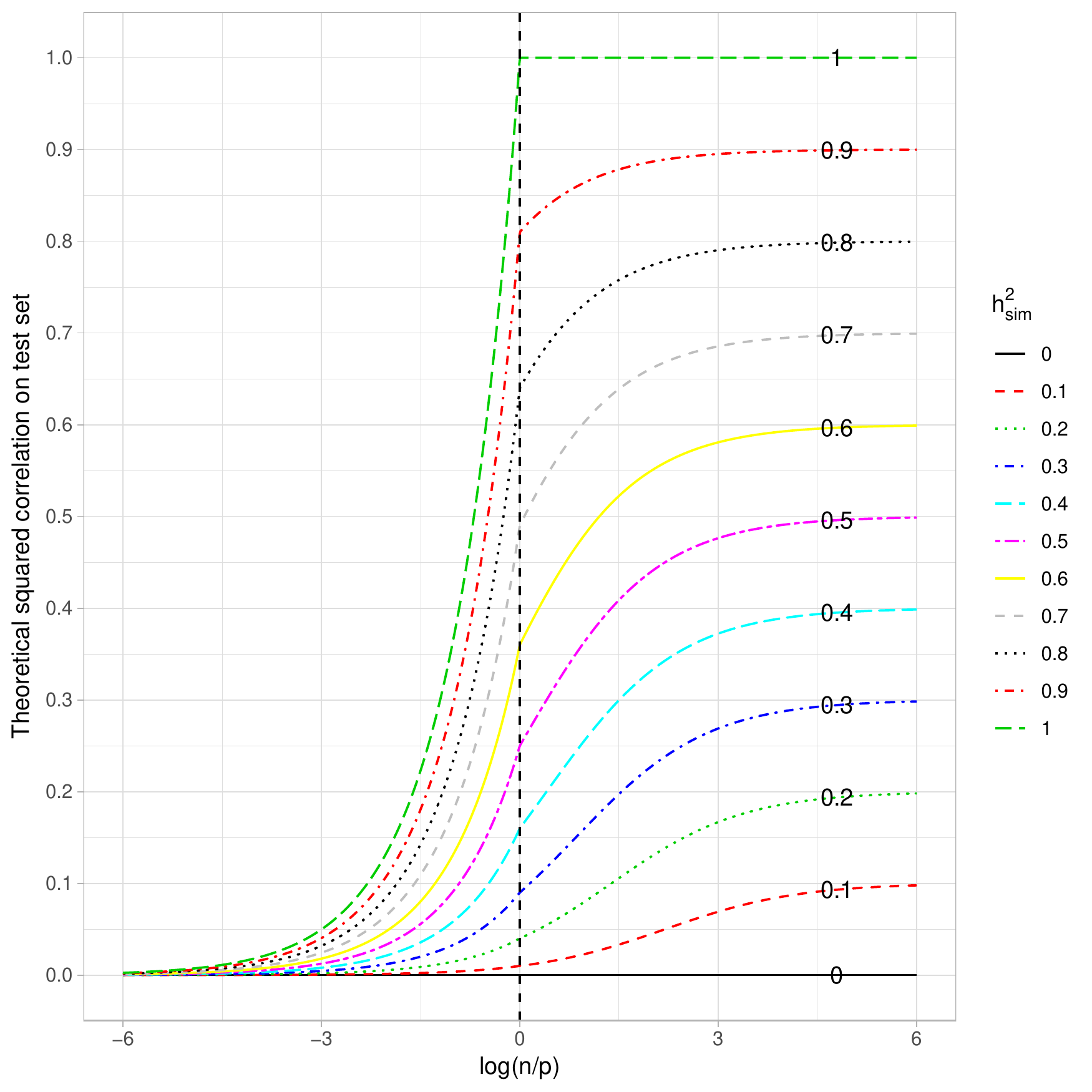}
    \caption{\label{fig:corr-theo}  Theoretical  squared correlation between phenotype and its prediction  with respect to  the log ratio of the number of individuals over the number of variants in the training set. Each curve corresponds to a given heritability (in the narrow sense).}
\end{figure}

When considering this  theoretical squared correlation with respect to  the log ratio of the number of individuals over the number of variants in the training set (Figure \ref{fig:corr-theo}), we  have, as expected, an increasing function. Similarly, the higher the heritability, the higher the squared correlation. We also observe that when $n\gg p$, the squared correlation tends toward the simulated heritability. Conversely, when $p \gg n$, it is close to zero whatever the heritability.

%-----------------------
\subsection{Simulations and real data\label{sec:simulations}}

 Since  narrow-sense heritability is a quantity  that relates  to a model, we
will  first illustrate our contributions via simulations where the true model is
known.   We perform two different types of simulation: fully synthetic  simulations  where
both genotypes and phenotypes are drawn from statistical distributions,  and semi-synthetic
simulations where  UK Biobank genotypes  are used to simulate phenotypes.
We also illustrate our contributions  using height and body mass index  (BMI)  from the UK Biobank  dataset.

\bigskip

We first assess the performance of GCV
for heritability estimation and then  look at  the accuracy of the
prediction when the ratio of the number of individuals  to  the number of variants varies in the
training set.

% Three types of data: simulated, semi-suynthetic and UKBB

\subsubsection{ UK Biobank  dataset}
\label{UKBB_filters}
%DIF < UKbiobank is a large open-access project pulling genotyping data from the British population.
The present analyses were conducted under UK Biobank data application number
45408. The UK Biobank  dataset consists of  $\simeq$ 784K autosomal SNPs describing $\simeq$ 488K individuals. We applied relatively stringent quality control and minor allele frequency filters to the dataset (callrate for individuals and variants $>$ 0.99 ; p-values of Hardy-Weinberg equilibrium test $>$ 1e-7 ; Minor Allele Frequency $>$ 0.01), leading to  473054 and 417106 remaining individuals and SNPs respectively.

\bigskip

Two phenotypes were considered in our analyses:  height (standing) and BMI.  In order to have a homogeneous population for the analysis of these real phenotypes, we retained only those  individuals who had reported their ethnicity as white British and whose Principal Component Analysis (PCA) results obtained by  UK Biobank were consistent  with their self-declared ethnicity . In addition, each time we subsampled individuals   we  removed related individuals (one individual  in all pairs  with a Genetic Relatedness Matrix (GRM) coefficient $>$ 0.025 was removed, as in
\citet{yang2011genome}. Several covariates were also considered in the analysis of these phenotypes: the sex, the year of birth, the recruitment center, the genotyping array,  and the first 10 principal components computed by  UK Biobank.

\subsubsection{Synthetic genotype data \label{sec:syntheticdata}}

The synthetic genotype matrices are simulated as in 
\citet{golan_measuring_2014} 
and
\citet{de_vlaming_current_2015}. This corresponds to a scenario with independent loci or perfect linkage equilibrium.
%DIF < The real genotype matrices are extracted from the UKBiobank dataset where loci are in linkage disequilibrium. These scenarios are respectively referred to as "fully-simulated" and "data-based simulated".  Details of the simulations are given below.
%DIF < A real data set from UKBiobank  with the height as phenotype is  also considered.
%DIF > The real genotype matrices are extracted from the UK Biobank dataset where loci are in linkage disequilibrium. These scenarios are respectively referred to as "fully-simulated" and "data-based simulated".  Details of the simulations are given below.
%DIF > A real data set from UK Biobank  with the height as phenotype is  also considered.

\bigskip

To simulate synthetic genotypes for $p$ variants, we first set a vector of variant frequencies $f \in \mathbb{R}^p$, with these frequencies  independently following  a uniform distribution $\mathcal{U} \left( [ 0.05, 0.5 ] \right)$. Individual genotypes are then drawn from  binomial distributions with proportions $f$, to form the genotype matrix $\textbf{M}$. A matrix of standardized genotypes $\textbf{Z}^*$ can be obtained by standardizing $\textbf{M}$ with the true variant frequencies $f$.

%We removed individuals with missing height, sex and year of birth.

\subsubsection{Simulations to assess heritability estimation using GCV}

We consider both synthetic and real genetic data, and simulate associated phenotypes.
%We then use GCV for  estimating $\lambda$ via the two above described solutions and derive $h^2_g$ using the link between $h^2_g$ and $\lambda$ described in Equation \ref{eq:lambda_to_h2g}.
%In a second part, we asses the alternative estimation of heritability estimation using $h^2_{r_1}$ and  $h^2_{r_2}$ (see Equations \ref{eq:train-test-heritability}), the ratios computed  from train and test sets.

%Having guarantees on the reliability of GCV we next invest heritability estimation using $h^2_{r_1}$ and $h^2_{r_2}$.

\bigskip

In  the two simulation scenarios  we investigate the influence  on heritability estimation of the following three parameters : the shape of the genotype matrix in the training set (the ratio between $n$ the number of individuals and $p$ the number of variants),  the fraction of variants with causal effects $f_c$,  and the true heritability $h^2_{sim}$. The tested levels of  these quantities are shown  in Table \ref{tab:table_levels_parametres_simulation}.

\begin{tableth}
    \begin{tabular}{c c}
    	\hline
         Parameters & Levels \\
         \hline
         n/p & Simulation : 1000/10000 ; 5000/100000 ; 10000/500000 \\
         & Data-based  : 1000/10000 ; 5000/100000 ; 10000/417106 \\
         $f_c$ & 0.1 ; 0.5 ; 1 \\
         $h^2_{sim}$ & $\lbrace 0.1, ..., 0.9 \rbrace$ \\
    \end{tabular}
    \caption{Table of  the parameters sets  of the  simulations. $n/p$ : the ratio of the dimensions of the genotype matrix. $f_c$ : proportion of causal variants. $h^2_{sim}$ :  simulated heritability.}
    \label{tab:table_levels_parametres_simulation}
\end{tableth}

\bigskip
%For the "fully simulated" scenario we first set a vector of allelic frequencies $\textbf{f} \in \mathbb{R}^p$ with variant frequencies following independently a uniform distribution $\mathcal{U} \left( [ 0.05, 0.5 ] \right)$. Individual genotypes are  drawn from  binomial distributions with proportions $\boldsymbol{f}$.
%Let denote by $\boldsymbol{M}$ the matrix of genotypes and $\boldsymbol{Z}$ its standardized version.
%$\forall i \in \llbracket 1, n \rrbracket, ~ M^T_i \sim \mathbb{B} \left( \textbf{f}, 2\right)$.

For each simulation scenario and for a given a set of parameters ($n$,$~p$,$~f_c$,$~h^2_{sim}$), the simulation of the phenotype starts with a matrix of standardized genotypes (either a synthetic genotype matrix $\textbf{Z}^*$ standardized with the true allele frequencies,  as described in Section \ref{sec:syntheticdata},  or a matrix of empirically standardized genotypes $\textbf{Z}$ obtained from  UK Biobank  data). To create the vector of genotype effects $u$,  $p \times f_c$ causal SNPs are randomly sampled and their effects are sampled from a multivariate normal distribution  with  zero mean and a covariance matrix  $\tau \textbf{I}_{p \times f_c}$ (where  $\tau = \frac{h^2_{sim}}{p \times f_c}$), while the remaining $p \times\left( 1 - f_c \right) $ effects are set to 0. The vector of environmental effects  $\textbf{e}$ is sampled from a multivariate normal distribution with zero mean and a covariance martrix $\sigma^2 \textbf{I}_n$, where   $\sigma^2 = 1 - h^2_{sim} $. The phenotypes are then generated as $\textbf{y} = \textbf{Z}^* u + \textbf{e}$ and $\textbf{y} = \textbf{Z} u + \textbf{e}$, for the fully synthetic scenario and the semi-synthetic scenario respectively. A standardization set of 1000 individuals (that will be used for the GCV approach based on two datasets) is also generated for each scenario  in the same way.

%DIF < UKbiobank is a large open-access project pulling genotyping data from the British population. The considered dataset consists of  $\simeq$ 784K SNPs describing $\simeq$ 488K patients. Relatively stringent quality control procedures were applied to the dataset ( callrate $>$ 0.99 ; Minor Allele Frequency $>$ 0.05 ; p-values of Hardy-Weinberg equilibrium test $>$ 1e-6 ), leading to 231019 remaining SNPs. We removed individuals with missing height, sex and year of birth. 
%DIF > UK Biobank is a large open-access project pulling genotyping data from the British population. The considered dataset consists of  $\simeq$ 784K SNPs describing $\simeq$ 488K patients. Relatively stringent quality control procedures were applied to the dataset ( callrate $>$ 0.99 ; Minor Allele Frequency $>$ 0.05 ; p-values of Hardy-Weinberg equilibrium test $>$ 1e-6 ), leading to 231019 remaining SNPs. We removed individuals with missing height, sex and year of birth.
%For the "data-based simulated" scenario we randomly chose $n$ individuals. The $p$ variants are taken in original order to keep LD structure.

%In practice the simulation of large scale matrices can be extremely time consuming. We applied the same idea as \cite{de_vlaming_current_2015} to speed up the simulations. We simulated an $(n_{max} = 10 000$ $\times$ $p_{max} = 500 000)$ matrix and used it for all the 27 $\times$ 3 ( $h^2_{sim} \times f_c$ ) scenarios. Studies of smaller size matrices were realized from a subset of this big matrix.

\bigskip

Applying GCV to  large-scale  matrices can be extremely  time-consuming, since it requires  the computation of the GRM associated with $\textbf{Z}^*$ or $\textbf{Z}$ and the eigen decomposition of the GRM. For this reason we employed the same strategy as \mbox{%DIFAUXCMD
\citet{de_vlaming_current_2015} }\hspace{0pt}%DIFAUXCMD
in order  to speed up both simulations and analyses  by making it possible to test more than one combination  of simulation parameters . We simulated an $(n_{max} = 10 000$ $\times$ $p_{max} = 500 000)$ genotype matrix for the training set  in  the fully synthetic scenario and used  this simulated matrix  for all the 9 $\times$ 3 $\times$ 3 = 81 ($h^2_{sim} \times f_c \times n/p$) parameter combinations. Similarly, we sampled $n_{max} = 10000$ individuals from the  UK Biobank dataset to obtain  an $(n_{max} = 10 000$ $\times$ $p_{max} = 417106)$ genotype matrix for the training set  in  the semi-synthetic scenario. Smaller  matrices were then  created  from a subset of these two large matrices (note that for subsets of the real genotype matrix  we took variants in the original order to  keep the  linkage disequilibrium structure).  Consequently,  computation of the GRM and its eigen decomposition  needed  to be performed only once for each $n/p$ ratio considered.
%For each parameter combination, we simulated a vector of genetic effects $u$ and an environmental component vector $\textbf{e}$.

\bigskip

The  fully synthetic and the semi-synthetic scenarios were each  replicated 30 times.

\subsubsection{Simulations to assess prediction accuracy}

 We performed fully synthetic simulations for different ratios $\frac{n}{p}$ in order  to study the behavior of the mean
prediction error  and  the correlation between the phenotype and its prediction . We considered
a training set of size $n = 1000$, and a test set of size
$n_{te} = 5000$.  The maximum number of variants was
set to $p_{max}= 50000$ and the heritability to $h^2 = 0.6$. We first
simulated a global allelic frequency vector
$f \sim \mathcal{U}_{p_{max}} (0.05,0.5)$  and a global vector of
genetic effects $ u \sim \mathcal{N} \left( 0_{p_{max}}, \frac{h^2}{p_{max}}
  \textbf{I}_{p_{max}} \right).$

\bigskip

For each subset of variants of size $p < p_{max}$, we selected a vector of genetic effects composed of
the $p$ first components of $ u$ multiplied  by a
$\sqrt{\frac{p_{max}}{p}}$ factor assuring a total variance of 1 and
  $\text{var} (u^p) = \frac{h^2}{p} \textbf{I}_p$.:
  $ u^p = ( u_1, ..., u_p ) \times
  \sqrt{\frac{p_{max}}{p}}$.
The genotype matrix $\textbf{M}_{te}$ was then simulated  and its normalized version  $\textbf{Z}_{te}^*$ computed as described in Section \ref{sec:syntheticdata}. The normalization used the first $p$ components of $f$.
The noise vector
$\textbf{e}_{te} \sim \mathcal{N} ( \textbf{0}_{n_{te}}, ( 1 - h^2 ) \textbf{I}_{n_{te}} )$ and a vector of phenotypes  $\textbf{y}_{te} = \textbf{Z}_{te}^* u^p + \textbf{e}_{te}$
were eventually simulated. 

%$\lambda_{opt} = p \frac{1 - h^2}{h^2}$.

\bigskip

We generated 300 training sets by simulating the  normalized genotype matrix, noise and phenotype using the same process as for the test set.   Here,  the training set index is denoted as $k$ . A prediction  $\hat{\textbf{y}}_{te,k}$ for the test set was made with each training set  using the ridge estimator of $\boldsymbol u^p$ obtained with $\lambda = p \frac{1 - h^2}{h^2}$, and the following empirical quantities  were estimated:
 $\text{err}_p =  \frac{1}{300} \sum_{k} \frac{1}{n_{te}} \left\Vert \textbf{y}_{te,k} - \hat{\textbf{g}}_{p} \right\Vert^2_2$,
$\text{bias}_p^2 = \frac{1}{n_{te}} \sum_{i \in \llbracket 1,n_{te} \rrbracket} \left( \left[ \textbf{Z}_{te} u^p  - \hat{\textbf{g}}_{p} \right]_i \right)^2$
and
$\text{var}_p =  \frac{1}{300} \sum_{k} \frac{1}{n_{te}} \left\Vert \hat{\textbf{y}}_{te,k} - \hat{\textbf{g}}_{p} \right\Vert^2_2$, where $\hat{\textbf{g}}_{p} = \left( \frac{1}{300} \sum_{k \in \llbracket 1, 300 \rrbracket} \left[ \hat{\textbf{y}}_{te,k} \right]_i \right)_{i \in \llbracket 1,n_{te} \rrbracket}$. The squared correlation between $\hat{\textbf{y}}_{te,k}$ and $\textbf{y}_{te,k}$
was also estimated.

\bigskip

We considered the following numbers of variants:
\begin{align*}
  p        \in \lbrace & 50000, 25000, 16667, 12500, 10000, 5000, 3333, 2500, 2000, 1667,\\
                       & 1429, 1250, 1111, 1000, 500, 136, 79, 56, 43, 35, 29, 25, 22, 20 \rbrace .
\end{align*}

%-----------------------
\subsection{Prediction of Height and BMI using  UK Biobank  data}

To experiment on UK Biobank for assessing the prediction accuracy,  for each phenotype we considered three sets of data: a training set  for the purpose of  learning genetic effects, a standardization set for learning non-penalized effects (covariates and intercept), and  a test set  for assessing  predictive power. Pre-treatment filters (as described in section \ref{UKBB_filters}) were systematically applied on the training set. We computed the estimation of genetic effects using the projection-based approach to take into account non-penalized effects, where the penalty parameter  was  obtained by GCV with the same projection approach:
\[ \hat{u}_{R} = \textbf{Z}_{tr}^T \textbf{C}_{tr}^T \left( \textbf{C}_{tr} \textbf{Z}_{tr} \textbf{Z}_{tr}^T \textbf{C}_{tr}^T + \hat{\lambda}_{GCV} \textbf{I}_{n-r} \right)^{-1} \textbf{C}_{tr} \textbf{y}_{tr}  . \]
   
\bigskip   
   
We then estimated non-penalized effects (here $\textbf{X}$ contains the intercept):
\begin{align}
    \hat{\beta} &= \left( \textbf{X}_{std}^T \textbf{X}_{std} \right)^{-1} \textbf{X}_{std}^T \left( \textbf{y}_{std}\right).
\end{align}

\bigskip

Finally, we applied these  estimations on the  test set:
\begin{align*}
    \hat{\textbf{g}}_{te} &= \textbf{Z}_{te} \hat{u}_{R}, \\
    \hat{\textbf{f}}_{te} &= \textbf{X}_{te} \hat{\beta}, \\
    \Tilde{\textbf{y}}_{te} &= \textbf{y}_{te} - \hat{\textbf{f}}_{te},
\end{align*}
in order to compute the Mean Square Error  = $ \frac{1}{n_{te}} ( \tilde{\textbf{y}}_{te} - \hat{\textbf{g}}_{te} )^T ( \tilde{\textbf{y}}_{te} - \hat{\textbf{g}}_{te} )$ between the phenotype residuals $\Tilde{\textbf{y}}_{te}$ after removal of non-penalized effects and $\hat{\textbf{g}}_{te}$.

\bigskip

This procedure was performed for different ratios $\frac{n}{p}$  using different sized  subsets of individuals  for the training set, while keeping all the variants that passed pre-treatment filters (see Table \ref{tab:table_n_pouvoir_pred_ukbb}).

\begin{tableth}
    \begin{tabular}{c c}
        \hline
         Set & Size \\
        \hline
         Training & $\lbrace 1000, 2000, 5000, 10000, 20000 \rbrace$ \\
         Standardization & 1000 \\
         Test & 1000
    \end{tabular}
    \caption[Size of training, standardization and test sets for the evaluation of predictive power on real data.] {Size (number of individuals) of training, standardization and test sets for  assessing  predictive power on real data.}
    \label{tab:table_n_pouvoir_pred_ukbb}
\end{tableth}

%\begin{tableth}
%    \begin{tabular}{|c||c c c c c |}
%         \hline
%         \textbf{Taille de l'ensemble d'apprentissage} & 1000 & 2000 & %5000 & 10 000 & 20 000 \\
%         \hline
%         \textbf{Nombre de répétitions} & 100 & 70 & 50 & 20 & 10  \\
%         \hline
%    \end{tabular}
%    \caption[Nombre de répétions pour l'évaluation du pouvoir prédictif sur données réelles.]{Nombre de répétions pour l'évaluation du pouvoir prédictif sur données réelles. Valeurs à vérifier}
%    \label{tab:nbr_repet_pouvoir_pred_ukbb}
%\end{tableth}

\bigskip

For each number $n$ of individuals considered in the training set, the sampling of these individuals was repeated several times,  as seen in Table \ref{tab:nbr_repet_pouvoir_pred_ukbb}, in order to account for the variance of the estimated genetic effects due to sampling.

\begin{tableth}
    \begin{tabular}{c c c c c c }
         Size of the training set & 1000 & 2000 & 5000 & 10 000 & 20 000 \\
         \hline
         Number of repetitions & 100 & 70 & 50 & 20 & 10  \\
    \end{tabular}
    \caption[Number of repetitions for assessing predictive power on real data.]{Number of repetitions for the evaluation of the predictive power on real data.}
    \label{tab:nbr_repet_pouvoir_pred_ukbb}
\end{tableth}

%%%%%%%%%%%%%%%%%%%%%%%%%%%%%%%%%%%%%%%%%%%%%%%
\section{Results}
%%%%%%%%%%%%%%%%%%%%%%%%%%%%%%%%%%%%%%%%%%%%%%%

%-----------------------
\subsection{Generalized Cross Validation for heritability estimation}
\subsubsection{Simulation results}

%Idée de présentation : on choisit bien $\lambda$ ( et donc on peut bien estimer héritabilité ) donc on sait qu'on fait la meilleur prédiction possible. Mais en fait ces prédictions elles sont mauvaises, grâce à l'$h^2$ test.

 For the two simulation scenarios we  look at the difference between the estimation of $h^2_g$ by GCV and the simulated heritability $h^2_{sim}$ in different configurations of study size $n/p$, $h^2_{sim}$ and the  fraction of causal variants $f_c$ . Similarly,  we look at the difference between the estimation by the classical mixed model approach and  the  simulated heritability.
 In our simulations $f_c$ was seen to have no influence, and so only  the influence of the remaining parameters is shown  in Figure \ref{grid_boxplot_ecart_h2g_h2sim_fc_0.1}.  For full results see Supplementary Material.

\begin{figure}
\begin{tabular}{c c}       \includegraphics[width=0.48\linewidth]{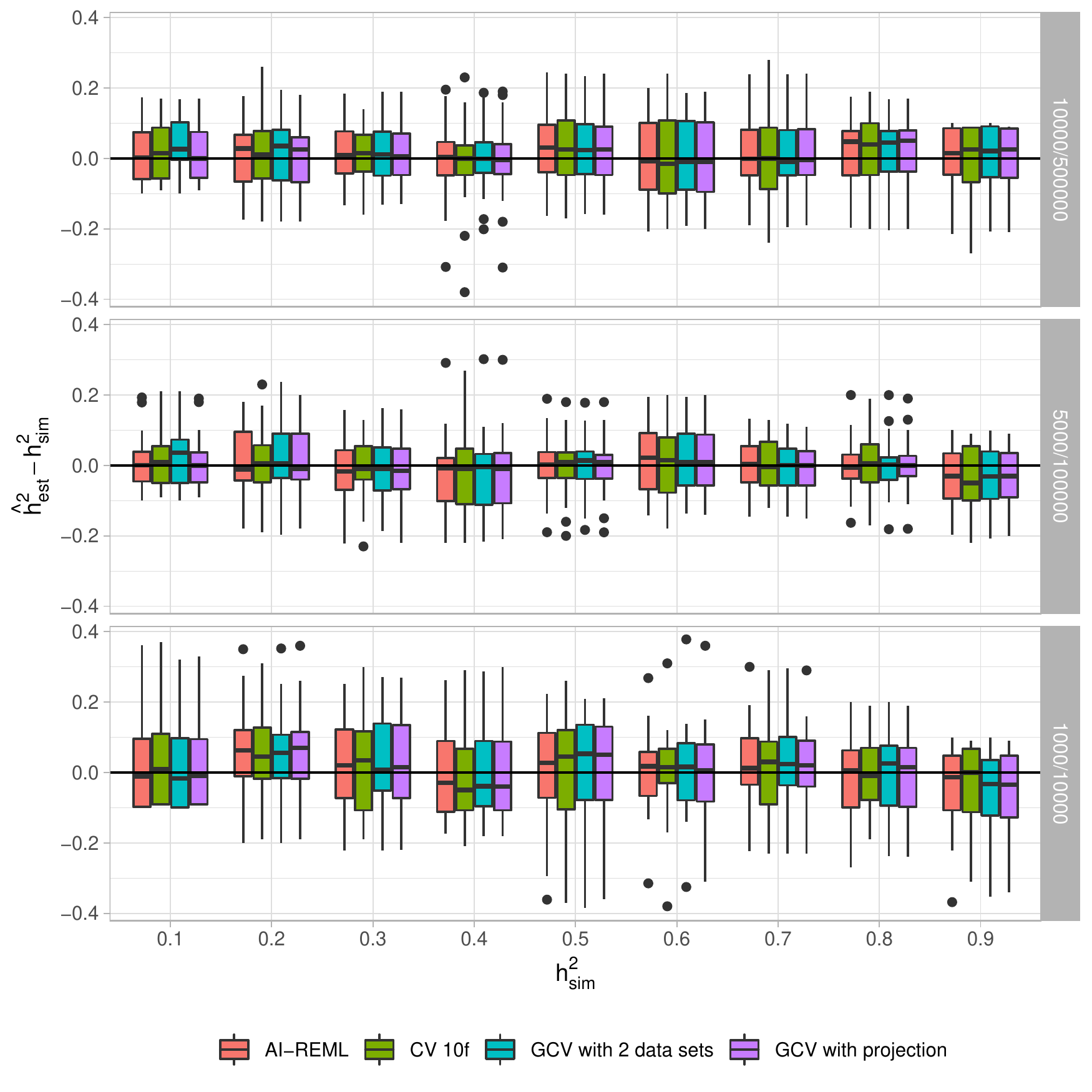} &      \includegraphics[width=0.48\linewidth]{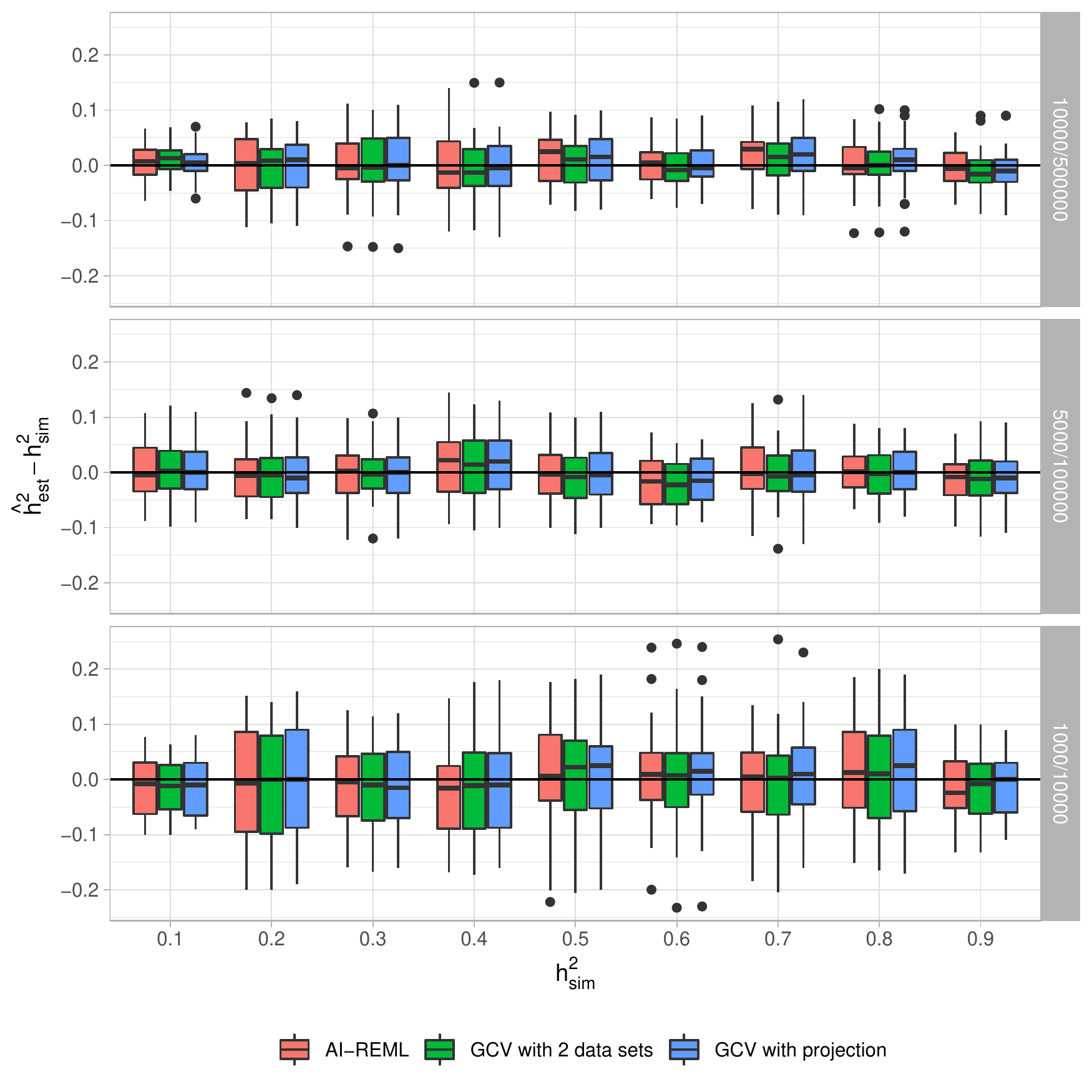} \\
     \parbox{0.5\textwidth}{(A) {Fully-synthetic simulation scenario: variants are simulated independently.}} &
       \parbox{0.5\textwidth}{ (B) {Semi-synthetic simulation scenario: correlation between variants.}}
\end{tabular}
    \caption{Distribution of $( h^2_{est} - h^2_{sim} )$ for different parameter combinations with 30 replications. Panel (A) corresponds to data simulated under the  ``fully synthetic''  procedure, while panel (B) corresponds to the  ``semi-synthetic simulation'' procedure. Each sub-panel corresponds to a different value of $n/p$. In both scenarios 10$\%$ of the variants have causal effects. For each panel, the horizontal axis corresponds to the simulated heritability $h^2_{sim} \in \lbrace 0.1, ...,0.9\rbrace$ and the vertical axis corresponds to $(h^2_{est} - h^2_{sim})$. Heritability estimations are done with the random effects model using AI-REML and with ridge regression using 3 approaches for the choice of $\lambda$ : GCV with a projection correction, GCV with a two-dataset correction  and a 10-fold cross-validation (CV 10f).}
    \label{grid_boxplot_ecart_h2g_h2sim_fc_0.1}
\end{figure}

\bigskip

%The first result we want to check is our ability to correctly estimate $\lambda$ and thus $h^2_g$.
For the fully-simulated scenario, the two GCV approaches give very similar results and  appear  to provide an unbiased estimator of $h^2$. They compare very well with the estimation of heritability by ridge regression with a 10-fold CV. Moreover, the variance of the GCV estimators does not  appear  higher than the  variance  of 10-fold CV . Our  choice of using GCV  in place of a classical CV approach   for estimating  heritability by ridge regression is therefore validated. 

\bigskip

In the case of the  semi-synthetic simulations,  here too  both GCV approaches provide a  satisfactory  heritability estimation.

\bigskip

For both simulation scenarios  we also note  that the classical mixed model approach (using the AI-REML method in the \texttt{gaston} R package) gives heritability estimations that are very similar  to those obtained  using  the GCV approaches.
%For each method and each $h^2_{sim}$ the mean of $h^2_g$ is close to expected value.
The value of simulated heritability does not  appear  to have a strong effect on the quality of the  heritability estimation.
%The fraction of causal variant does not seem to have much effects on the estimation.
On the other hand, the ratio $n/p$ seems to have a real impact on estimation variance,  with lower ratios leading to lower variances, which initially might appear surprising. One possible explanation for this is that  in our simulations $n$ increases as the ratio $n/p$ decreases. 
\citet{visscher_general_2015} 
showed that the variance of the heritability is a decreasing function of $n$,  which could explain the observed behaviour.

%Similar results are observed in the "data-based" simulation \ref{grid_boxplot_avec_LD_h2g}. GCV provided satisfying heritability estimation for both correction and any combination of $h^2_{sim}$, $f_c$ or $n/p$. Again the fraction of causal variant does not have a impact on the simulation and the smaller the $n/p$ ratio the smaller the variance of the estimation.

%For both simulation design, ridge regression using GCV provided results close to the one from random effects model. This is not a surprise but it does show the use of ridge regression for heritabilty estimation.

\bigskip

\subsubsection{Illustration on UK Biobank}

We now compare heritability estimations between the two GCV approaches and the classical mixed model approach  for height and BMI, on a training set of 10000 randomly sampled individuals (the training set being  of the same size as for the  simulated data).  All three approaches take account of covariates and  the intercept. The AI-REML approach also  uses  a projection matrix to deal with covariates. For the GCV approach based on two datasets, a standardization set of 1000 individuals is also sampled, and for comparison purposes we have chosen  to apply this two-set strategy to the classical mixed model approach as well.

\bigskip

Since the true heritability is of course unknown with real data, the sampling of the training and standardization sets is repeated 10 times in order to account for heritability estimation variability.  Note  that the SNP quality control and MAF filters were  repeated at each training set sampling and applied to the standardization set.

\bigskip

\begin{figure}
    \centering
    \includegraphics[width=0.7\linewidth]{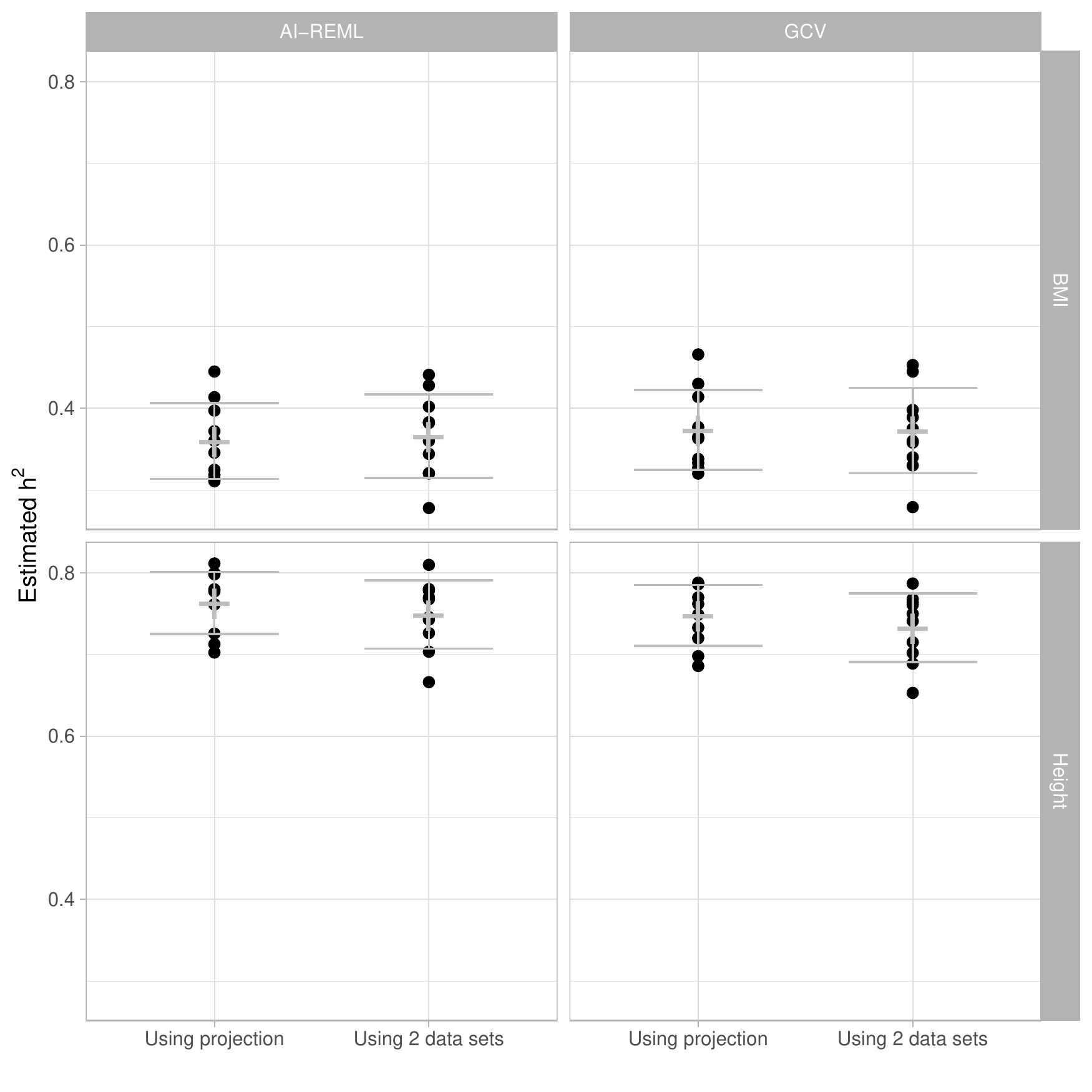}
    \caption{Heritability estimation of BMI and height using AI-REML and GCV, with the projection-based approach  andwith the two-set approach. We sub-sampled the original UK Biobank dataset 10 times for replication. The cross  corresponds  to the mean and the error bar to the mean +/- one standard deviation. }
    \label{GCV_UKBB}
\end{figure}

Figure \ref{GCV_UKBB} shows that  for each phenotype  the two GCV approaches and the classical mixed model approach (AI-REML) give similar estimations.  There is relatively little estimation variability, and any variability observed seems depend more on the individuals  sampled for the training set  than on the approach  used.

%-----------------------
\subsection{Prediction versus Heritability in the context of small additive effects}

\subsubsection{Prediction from synthetic data}

\begin{figure}
    \centering
    \begin{tabular}{c c}
        \includegraphics[width=0.48\linewidth]{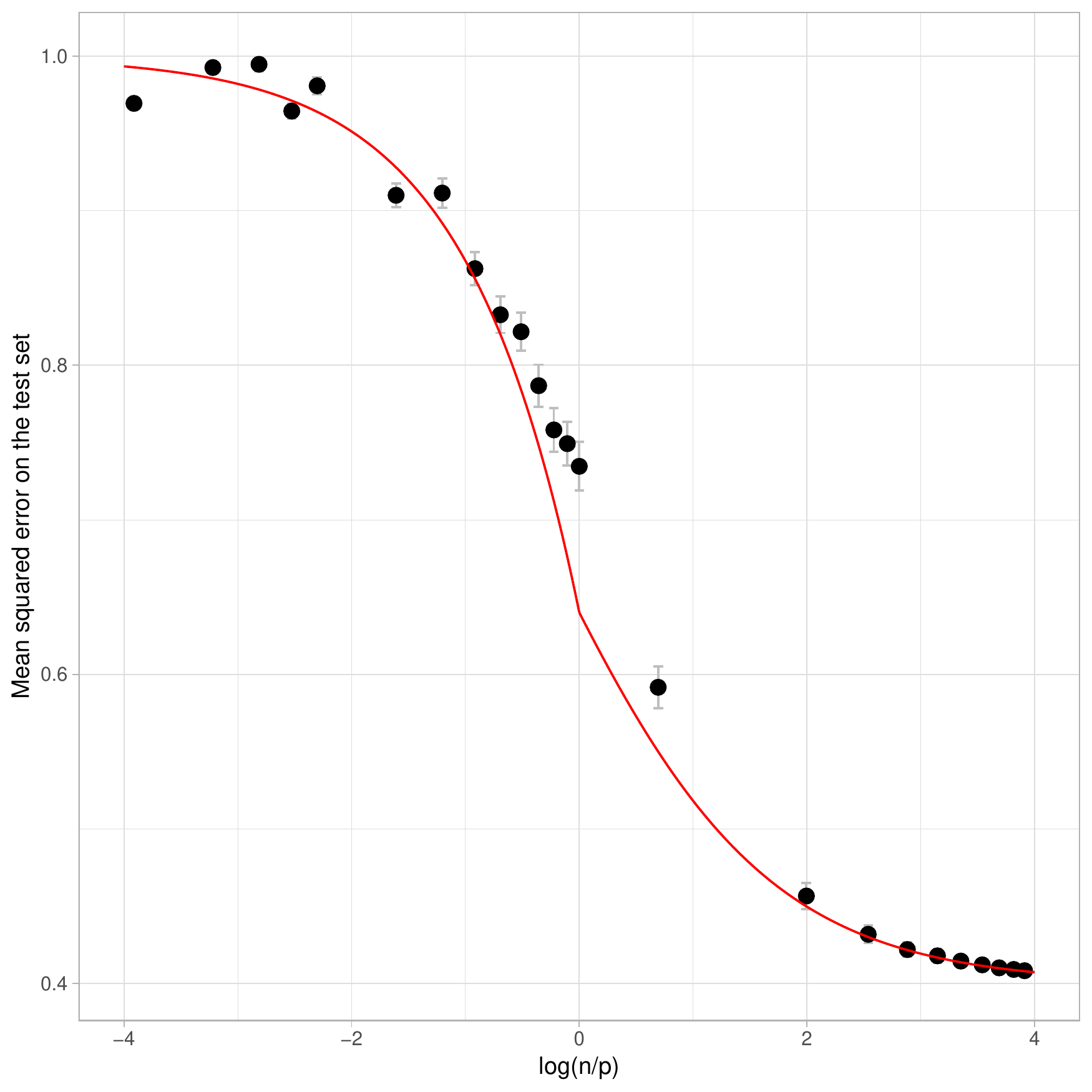} &
        \includegraphics[width=0.48\linewidth]{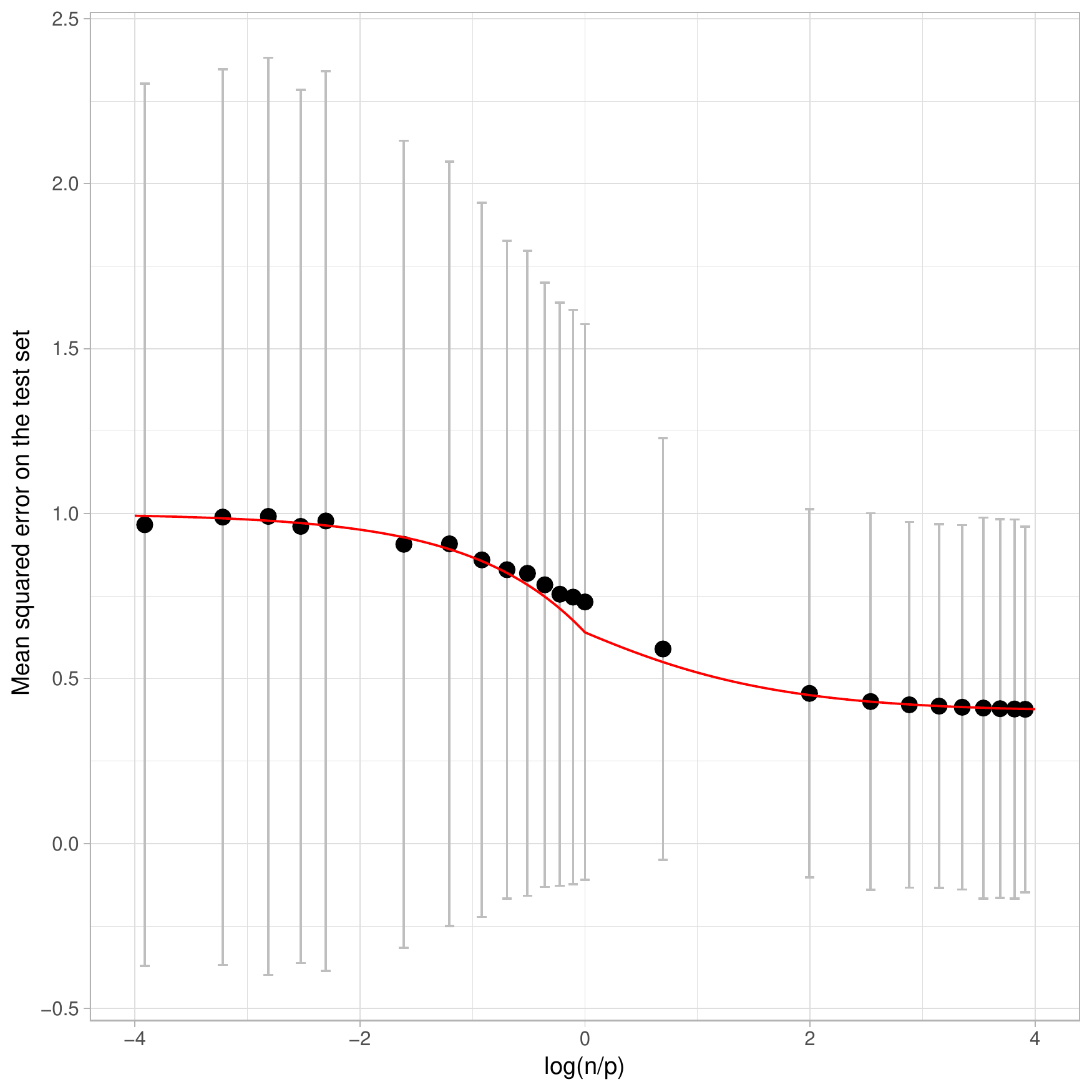} \\
        (A) & (B)
    \end{tabular}
    \caption{\label{fig:prediction-simu} Mean Squared Error of the prediction on the test set with respect to  $\log(\frac{n}{p})$, using  simulated data with $h^2=0.6$. The curves correspond to the theoretical  link for $h^2=0.6$. The black points correspond to the mean expectation for each value of $\log(\frac{n}{p})$ over 300 repetitions. The error bars in (A) and (B) correspond respectively to one standard deviation of the test set error obtained using two different aggregation strategies. On the left (A), we consider an aggregation strategy where each of the 300 training sets results in  a mean test set error, whereas on the right (B) each individual in  the test set results in an error averaged over all training sets.}
\end{figure}

As expected,  the mean of the test set error follows closely the  theoretical  curve when the $\log{\frac{n}{p}}$ varies (Figure \ref{fig:prediction-simu}).  When $n>p$,  the mean of the test set is close to the minimum possible error, which means that the ridge regression provides a reliable prediction on average.

\bigskip

Interestingly, if the mean error behaves as expected by our approximation, the standard deviation of the error may be very large.  Figures \ref{fig:prediction-simu}A and \ref{fig:prediction-simu}B  show  the same mean error with different error bars. Figure \ref{fig:prediction-simu}A plots the error bars corresponding to the training set variation: the mean test set error is computed for each training set and the error bars  show  one standard deviation across the 300 training sets.
Figure \ref{fig:prediction-simu}B plots the error bars corresponding to the variation of the errors across the test set.

\bigskip

The  error bars in   Figure \ref{fig:prediction-simu}B are much larger than  those in  Figure \ref{fig:prediction-simu}A,  which   shows that the variation in the prediction error  is mostly due to  the test individual whose  phenotype we  wish to predict, and depends little on  the training set. This may be explained by the fact that the environmental residual term can be very large for some individuals.  For these individuals the phenotype will be predicted with a  very large error even when $n\gg p$, that is to say  when the genetic term is correctly estimated, irrespective of  the training set (see  Supplementary Material).

\bigskip

\begin{figure}
    \centering
    \includegraphics[width=0.7\linewidth]{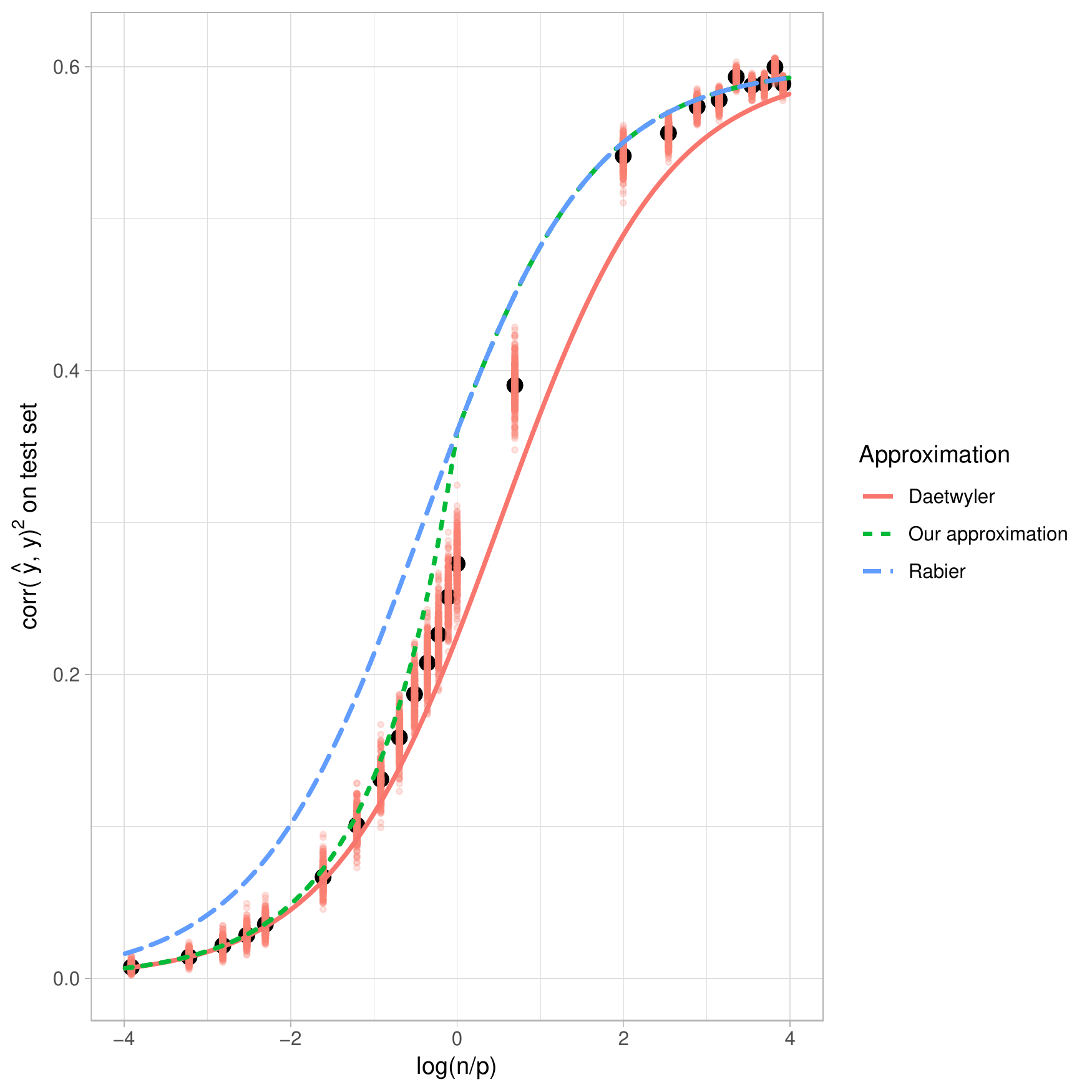}
    \caption{\label{fig:cor2-simu} Mean squared correlation between the phenotype and its prediction on the test set with respect to  $\log(\frac{n}{p})$, using simulated data with $h^2=0.6$. The salmon points correspond to the evaluation of the squared correlation and the black points correspond to the mean expectation for each value of $\log(\frac{n}{p})$ over 300 repetitions. Red dots correspond to training set replications. The red plain curve corresponds to Daetwyler's approximation for $h^2=0.6$, while the blue dashed curve corresponds to Rabier's approximation and the green dotted curve corresponds to ours.}
\end{figure}

The squared correlation between the phenotype and its prediction, as a function of $\log{\frac{n}{p}}$,  is also in line with our approximation  (Figure \ref{fig:cor2-simu}). As expected, when $n\gg p$, the squared correlation tends toward the simulated heritability. We compared our approximation with the  approximation  obtained by \citet{daetwyler_accuracy_2008} and observed that  although Daetwyler's  approximation is very similar to ours when  $p\gg n$, our simulation results make  Daetwyler's approximation  appear  under-optimistic when $n\gg p$ . Finally, we also compared our approximation with  that  obtained by \citet{rabier_accuracy_2016}, which is the same as ours when $n>p$. However, when $p>n$, Rabier's approximation  appears  over-optimistic.

\subsubsection{Prediction from  UK Biobank  data}
Let us consider the proposed   theoretical  approximation of the predictive power of ridge regression   with respect  to the $n/p$ ratio applied to  the  UK Biobank data, for height and BMI residuals (after removal of covariate effects and intercept).

\bigskip

The two phenotypes differ  considerably as regards  heritability: we estimate by the projection-based GCV approach that $73,33\%$ of height is ``heritable''  whereas only $33,91\%$ of BMI is ( on  average over the 10 training samples of 20000 individuals).

% Dealing with real data, we need to account for covariates in the model. We thus use the GCV approach with a projection matrix to estimate both heritability (as mentionned in section \ref{sec:}) and genetic effects $\boldsymbol u$, while accounting for covariates. For the estimation of the covariate effects, we use a standardization set of 1000 individuals. Finally, on the test set, we assess the power of the genetic term to predict the phenotype, after removal of the covariate effects.

% For each number $n_{tr}$ of individuals considered in the training set, the sampling of individuals was repeated 50 times (on average).

\bigskip

These estimated  values are close to  those  currently found in the literature \citep{ge_phenome-wide_2017}.
It is important to note that the  heritability estimation is strongly dependent on the filters.  Variations of  up to $20\%$  were observed in the estimations when  the filtering procedure setup was slightly modified .

\bigskip

A major difference between UK Biobank data and  our simulations  designed to check the proposed approximation  lies in the strong linkage disequilibrium present in the human genome.   Several  papers have proposed  using the  effective number of independent markers   to make adjustments in  the multiple testing framework \citep{li2012evaluating}, and we likewise
propose adjusting  our prediction model by taking into account an effective number of SNPs ($p_e$).
%DIF < Using the relation $\mathbb{E}_{\textbf{y}_{tr}, y_{te}, z_{te}} \left[ ( y_{te} - \hat{y}_{te} )^2 \right]= 1 - \frac{n_{tr}}{p} (h^2)^2$ when $p>n_{tr}$, we can use a simple linear regression to find the coefficient allowing to predict the real $\frac{n_{tr}}{p_e}$ ratio from the corresponding theoritical $\frac{n_{tr}}{p}$ ratio. 
 %DIF > Using the relation $\mathbb{E}_{\textbf{y}_{tr}, y_{te}, z_{te}} \left[ ( y_{te} - \hat{y}_{te} )^2 \right]= 1 - \frac{n_{tr}}{p} (h^2)^2$ when $p>n_{tr}$, we can use a simple linear regression to find the coefficient allowing to predict the real $\frac{n_{tr}}{p_e}$ ratio from the corresponding theoretical $\frac{n_{tr}}{p}$ ratio.
We  estimate the effective $\frac{n}{p_e}$ ratio for each training set and for each considered $n$ value using the  observed mean square errors, the estimated heritability, and the theoretical  relation in the case of independent variants
$\mathbb{E}_{\textbf{y}_{tr}, y_{te}, z_{te}} \left[ ( y_{te} - \hat{y}_{te} )^2 \right]= 1 - \frac{n}{p} (h^2)^2$ when $p>n$.
We then use a simple linear regression to find the coefficient between these estimated $\frac{n}{p_e}$ ratios and the corresponding real $\frac{n}{p}$ ratios.

\bigskip

Table \ref{table:effectivenumber} shows different but close effective numbers of SNPs for the two phenotypes.

\begin{tableth}
    \begin{tabular}{c c}
        \hline
        Phenotype     & $p/p_e$ \\
        \hline
        Height     & $5.01$   \\
        BMI        & $3.48$
    \end{tabular}
    \caption{Effective number of SNPs}
    \label{table:effectivenumber}
\end{tableth}

\bigskip

% TODO: Arthur tu peux mettre ce que tu as vraiment fait:

\begin{figure}
    \centering
    \begin{tabular}{c c}
        \includegraphics[width=0.48\linewidth]{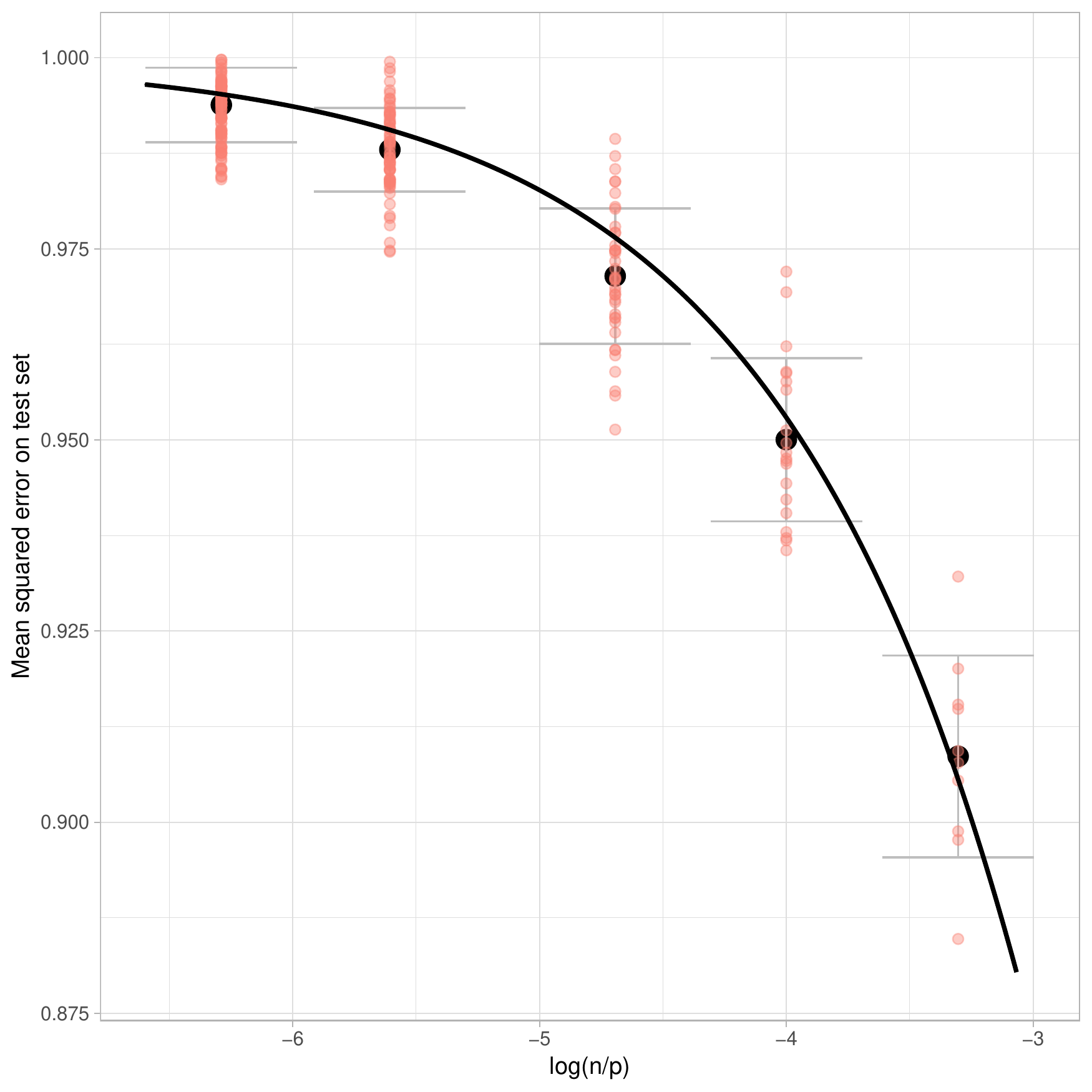} &
        \includegraphics[width=0.48\linewidth]{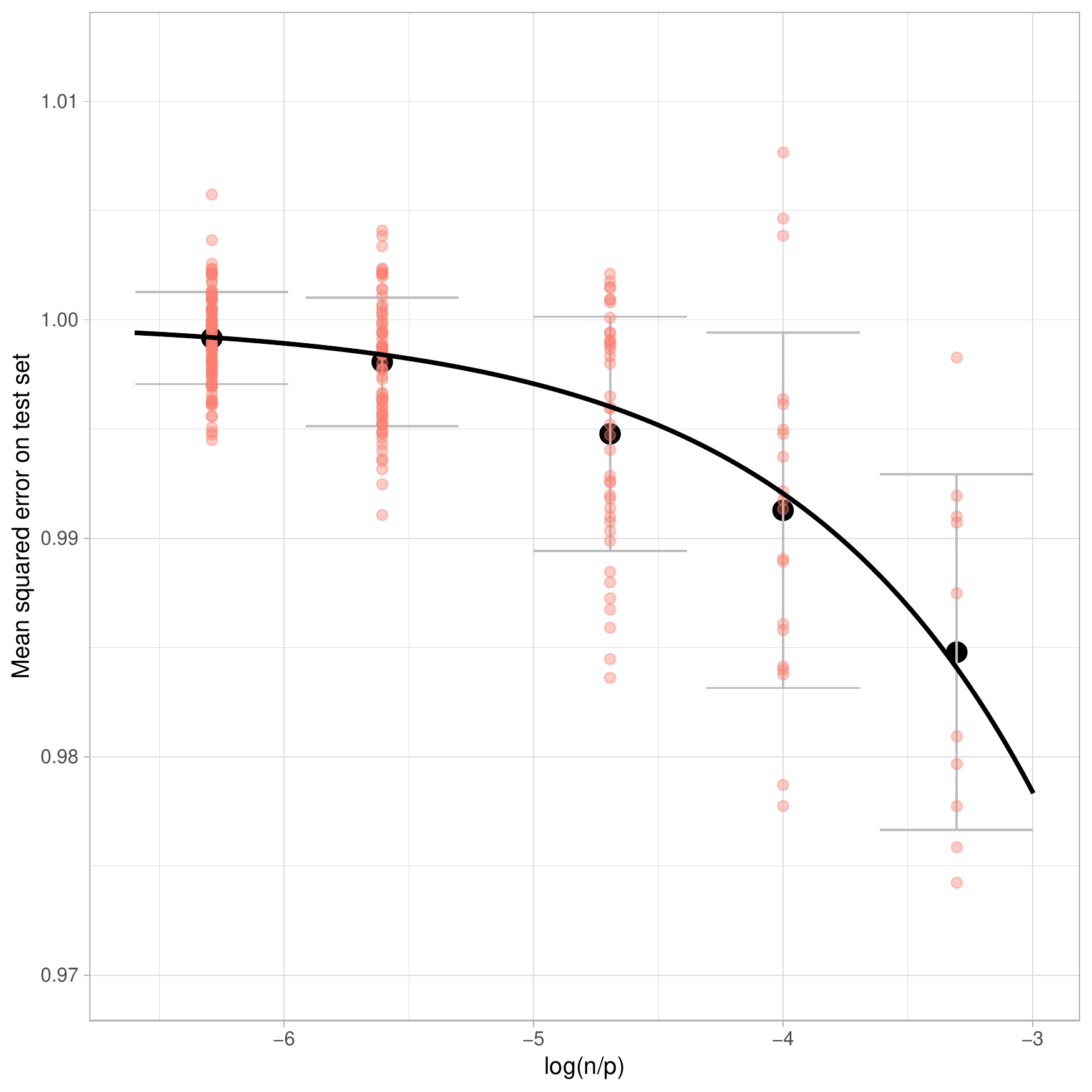} \\
        (A) Height & (B) BMI
    \end{tabular}
    \caption{Normalized Mean Squared Error on the test set for the prediction of Height (A) and BMI (B) with respect to  the log ratio of the number of individuals over the number of markers in the training set. Salmon dots correspond to training set replications, black dot to the mean of replications for different ratio and we show one standard deviation (over the training sets) of the mean test set error. The theoretical  curves are fitted using the estimated heritability and an effective  number of markers.}
    \label{pred_height_BMI}
\end{figure}

%We also consider normalizing the errors using the total variance of phenotype residuals (after removal of covariate effects and intercept).
We also consider normalizing the test set errors using the mean square error of phenotype residuals (after removing non-penalized effects).
Using  this error normalization and adjusting the theoretical curve for an  effective number of SNPs, we observe  a close fit between the estimated errors on the test set and their  theoretical values (Figure \ref{pred_height_BMI}).

%%%%%%%%%%%%%%%%%%%%%%%%%%%%%%%%%%%%%%%%%%%%%%%
\section{Discussion}
%%%%%%%%%%%%%%%%%%%%%%%%%%%%%%%%%%%%%%%%%%%%%%%

In this work  we investigated an alternative computation of genomic heritability  based on ridge regression. We proposed a fast,  reliable way to estimate the optimal penalisation parameter of the ridge via  Generalized Cross Validation  adapted for high dimension. The genomic heritability estimated from the GCV  gives  results comparable to mixed model AIREML estimates. It clearly demonstrates that a predictive criterion allows a reliable choice of the penalisation parameter and associated heritability, even when the prediction accuracy of ridge regression is low. Moreover, even  though  our approach does not formally consider  Linkage Disequilibrium, simulations showed that it still provides reliable genomic heritability estimates in presence of realistic  Linkage Disequilibrium .

\bigskip

We also provided  theoretical  approximations of the ridge regression prediction accuracy, in terms of both error and correlation between the phenotype and its prediction on new samples. These approximations perform well on synthetic data, in both  high and low dimensions. They rely on  the assumption that  individuals and markers are independent  in  approximating the empirical covariance matrices.  Our approximation of the prediction accuracy in terms of correlation proposes a good compromise between existing approximations. In particular,
it exhibits similar performances to \citet{daetwyler_accuracy_2008} when $p>n$ and to \citet{rabier_accuracy_2016} when $p<n$.

\bigskip

Our  theoretical  approximation of the prediction error is also  consistent  with the error observed on real genetic data when $p>n$, after adjusting for the effective number of independent markers. Unfortunately, due to  computational issues, we  were unable to  perform the analysis in the $n \simeq p$ case with real data.
However,  we observed that the prediction accuracy already reaches almost $ 15\%$ of the heritability of height when $n/p \simeq 5\%$, while \citet{de_los_campos_prediction_2013} suggested that its asymptotic upper bound is of the order of $20\%$ of the heritability because of incomplete LD between causal loci and genotyped markers.
Interestingly, ridge regression  is not affected by correlated predictors, and consequently it is not affected  by high LD between markers. When LD is  high, this  has the effect of reducing the degrees of freedom of the  model \citep{dijkstra2014ridge}, which results in an improved prediction accuracy  in comparison with a problem  having  the  same number of independent predictors and the same heritability.

\bigskip

Although  our approximations and simulation results tend to show that the prediction accuracy can reach the heritability value when $ n \gg p $, as already suggested by previous works \citep{daetwyler_accuracy_2008,rabier_accuracy_2016,de_vlaming_current_2015}, the large standard deviation of the prediction error that we observed between simulated individuals suggests that disease risk prediction from genetic data  alone  is not accurate at the individual level, even for a relatively high heritability value in the context of a  small additive effect hypothesis.

\bigskip

In direct continuity of this work, it would be interesting to investigate the behavior of prediction accuracy on real human data  where  $n \simeq  p$.  This would enable us to determine whether  our approximations still hold in that case, and even in the case where  $n>p$ (where we approximate the empirical covariance matrix of the markers to be diagonal) . It would show whether  it is possible for the prediction accuracy to  exceed  the upper bound proposed by \citet{de_los_campos_prediction_2013}. A further  prospect  would be to consider a  nonlinear  model extension via kernel ridge regression,  which may improve the prediction \citep{morota_kernel-based_2014}.

\bibliographystyle{frontiersinHLTH&FPHY} % for Health, Physics and Mathematics articles
\bibliography{biblio_article}

%%% Make sure to upload the bib file along with the tex file and PDF
%%% Please see the test.bib file for some examples of references

%%%%%%%%%%%%%%%%%%%%%%%%%%%%%%%%%%%%%%%%%%%%%
%%%%%% SUPPLEMENTARY 
%%%%%%%%%%%%%%%%%%%%%%%%%%%%%%%%%%%%%%%%%%%%%
\section*{Supplementary Material}

\subsection{A useful algebra for ridge regression}
\begin{align*}
    & \textbf{Z}^T \left( \textbf{Z} \textbf{Z}^T + \lambda \textbf{I}_p \right) = \left( \textbf{Z}^T \textbf{Z} + \lambda \textbf{I}_n \right) \textbf{Z}^T \\
   \Rightarrow & \left( \textbf{Z}^T \textbf{Z} + \lambda \textbf{I}_n \right)^{-1} \textbf{Z}^T \left( \textbf{Z} \textbf{Z}^T + \lambda \textbf{I}_p \right) \left( \textbf{Z} \textbf{Z}^T + \lambda \textbf{I}_p \right)^{-1} = \left( \textbf{Z}^T \textbf{Z} + \lambda \textbf{I}_n \right)^{-1} \left( \textbf{Z}^T \textbf{Z} + \lambda \textbf{I}_n \right) \textbf{Z}^T \left( \textbf{Z} \textbf{Z}^T + \lambda \textbf{I}_p \right)^{-1}\\
    \Rightarrow & \left( \textbf{Z}^T \textbf{Z} + \lambda \textbf{I}_n \right)^{-1} \textbf{Z}^T = \textbf{Z}^T \left( \textbf{Z} \textbf{Z}^T + \lambda \textbf{I}_p \right)^{-1}.
\end{align*}

\subsection{Computation of the GCV}

%We first present the computation of the LOO error then the computation of the GCV.

\subsubsection{Computation of the LOO error}

To compute the leave-one-out error ( LOO ) error, the  estimation of ridge regression parameters  without  individual $i$,  $\hat{u}_R^{-i}$, is required. Let us recall  the Sherman-Morrison-Woodbury's formula : let $\textbf{A} \in \mathcal{M}_p$ a non-singular matrix and $u,v \in \mathbb{R}^p$.

\begin{equation}
    \left( \textbf{A} + u v^T\right)^{-1} = \textbf{A}^{-1} - \frac{\textbf{A}^{-1} u v^T \textbf{A}^{-1}}{1 + v^T \textbf{A}^{-1} u}.
    \label{eq:formule_SMW}
\end{equation}

\bigskip

Using Sherman-Morrison-Woodbury's formula in the context of ridge regression, we have
\begin{align*}
    \left( \textbf{Z}_{-i}^T \textbf{Z}_{-i} + \lambda \textbf{I}_p \right)^{-1} &=  \left( \textbf{Z}^T \textbf{Z} - z_i z_i^T + \lambda \textbf{I}_p \right)^{-1} \\
    &= \left( \textbf{Z}^T \textbf{Z} + \lambda \textbf{I}_p \right)^{-1} + \frac{\left( \textbf{Z}^T \textbf{Z} + \lambda \textbf{I}_p \right)^{-1} z_i z_i^T \left( \textbf{Z}^T \textbf{Z} + \lambda \textbf{I}_p \right)^{-1} }{1 - z_i^T\left( \textbf{Z}^T \textbf{Z} + \lambda \textbf{I}_p \right)^{-1} z_i}
\end{align*}
with $z_i \in \mathbb{R}^p$ the column vector corresponding to the normalized genotypes of the $i$-th row (i.e. the $i$-th individual) of $\textbf{Z}$ and $\textbf{Z}_{-i} \in \mathcal{M}_{n-1,p} (\mathbb{R})$ the matrix $\textbf{Z}$ excluding its i-th row . 

\bigskip

Noticing that
\begin{align*}
    \textbf{Z}_{-i}^T \textbf{y}_{-i} = \textbf{Z}^T \textbf{y} - z_i y_i,
\end{align*}{}
it is straightforward to get
\begin{align*}
    \hat{u}_R^{-i} &= \left( \textbf{Z}_{-i}^T \textbf{Z}_{-i} + \lambda \textbf{I}_p \right)^{-1} \textbf{Z}_{-i}^T \textbf{y}_{-i} \\
    &= \left( \left( \textbf{Z}^T \textbf{Z} + \lambda \textbf{I}_p \right)^{-1} + \frac{\left( \textbf{Z}^T \textbf{Z} + \lambda \textbf{I}_p \right)^{-1} z_i z_i^T \left( \textbf{Z}^T \textbf{Z} + \lambda \textbf{I}_p \right)^{-1} }{1 - z_i^T\left( \textbf{Z}^T \textbf{Z} + \lambda \textbf{I}_p \right)^{-1} z_i} \right) \left( \textbf{Z}^T \textbf{y} - z_i y_i \right) \\
    &= \hat{u}_R - \left( \textbf{Z}^T \textbf{Z} + \lambda \textbf{I}_p \right)^{-1} z_i y_i + \frac{\left( \textbf{Z}^T \textbf{Z} + \lambda \textbf{I}_p \right)^{-1} z_i z_i^T \left( \textbf{Z}^T \textbf{Z} + \lambda \textbf{I}_p \right)^{-1} }{1 - z_i^T\left( \textbf{Z}^T \textbf{Z} + \lambda \textbf{I}_p \right)^{-1} z_i} \left( \textbf{Z}^T \textbf{y} - z_i y_i \right).
\end{align*}

\bigskip

Using that $ z_i^T\left( \textbf{Z}^T \textbf{Z} + \lambda \textbf{I}_p \right)^{-1} z_i = [h_{\lambda}]_{ii}$ and remembering that $y_i$ is a scalar we have
\begin{equation*}
    \hat{u}_R^{-i} = \hat{u}_R - \frac{ ( \textbf{Z}^T \textbf{Z} + \lambda \textbf{I}_p )^{-1} z_i (y_i - z_i^T \hat{u}_R ) }{1 - \left[ h_{\lambda} \right]_{ii}}.
\end{equation*}

\bigskip

Injecting this expression in the classic Mean Squared Error ( MSE ), the LOO error expresses as 
\begin{align}
    \text{err}^{LOO}(\lambda) &= \frac{1}{n} \sum_{i=1}^{n} \left( y_i - z_i^T \hat{u}_R^{-i}(\lambda) \right)^2 \\
    &= \frac{1}{n} \sum_{i=1}^{n} \left( \frac{y_i - z_i^T \hat{u}_R(\lambda)}{1 - \left[ h_{\lambda} \right]_{ii}} \right)^2 \label{eq:forme_LOO_pour_lisseur} \\
    & = \frac{1}{n} \textbf{y}^T ( \textbf{I}_n - \textbf{H}_{\lambda} ) ( \text{diag} (\textbf{I}_n - \textbf{H}_{\lambda}) )^{-2} ( \textbf{I}_n - \textbf{H}_{\lambda} ) \textbf{y},
    \label{eq:LOO_forme_matricielle}
\end{align}
where
$\textbf{H}_{\lambda} = \textbf{Z} \textbf{Z}^T ( \textbf{Z} \textbf{Z}^T + \lambda \textbf{I}_n )^{-1} = \textbf{Z} ( \textbf{Z}^T \textbf{Z} + \lambda \textbf{I}_p )^{-1} \textbf{Z}^T $ is the so-called  hat matrix because it transforms $\textbf{y}$
into $\hat{\textbf{y}}$:
\begin{align*}
    \hat{\textbf{y}} = \textbf{H}_{\lambda} \textbf{y} = \textbf{Z} \hat{u}_R(\lambda).
\end{align*}

\bigskip

An important point to notice is that this LOO is not a "true" n-fold cross validation because we standardize our data only once. For the more classical n-fold cross validation we would standardize each training set separately and use this standardization on the validation sample. While it may not look significant, we will show later that this unique standardization has important consequences. 

\subsubsection{Computation of the GCV error}
Generalized Cross validation ( GCV ) is an approximation of the LOO. First we introduce some notion about circulant matrices.

\bigskip

A matrix $\textbf{C}$ is called a circulant matrix if it is of the form
\[ \textbf{C} = \begin{pmatrix}
c_0 & c_1 & c_2 & \cdots & c_{n-1} \\
c_{n-1} & c_0 & c_1 & & c_{n-2} \\
c_{n-2} & c_{n-1} & c_0 & & c_{n-3} \\
\vdots & & & \ddots & \vdots \\
c_1 & c_2 & c_3 & \cdots & c_0
\end{pmatrix}\in \text{Circ}(n). \]
Such a matrix has constant diagonal coefficients. 
Let $\textbf{W}\in\mathcal{O}_n \left( \mathbb{C} \right)$ an orthogonal matrix such as $\left[ \textbf{W} \right]_{jk} = \frac{1}{\sqrt{n}} e ^{2 \pi i j k /n}$ with $j,k\in \left\lbrace 1,...,n \right\rbrace$. Then $\textbf{W}$ diagonalize all circulant matrices i.e.
\[\forall \textbf{C} \in \text{Circ}(n), ~ \exists \boldsymbol{\mathcal{D}} \in \mathbb{D}_n\left( \mathbb{C} \right) ~/~ \textbf{C} = \textbf{W} \boldsymbol{\mathcal{D}} \textbf{W}^*\]
with $^*$ the complex transpose operator.

\bigskip

The idea underlying GCV is to project the initial model in a well-chosen complex space such that the matrix $\textbf{H}_{\lambda} \in \text{Circ}(n)$. In this new model it is straightforward to compute the inverse of $\text{diag} (\textbf{I}_n - \textbf{H}_{\lambda})$ needed in  \eqref{eq:LOO_forme_matricielle}. This will shorten the computational time.

\bigskip

Let $\textbf{Z} = \textbf{U} \textbf{D} \textbf{V}^T$ be the singular value decomposition (SVD) of $\textbf{Z}$ with $\textbf{U} \in \mathcal{O}_n \left( \mathbb{R} \right)$, $\textbf{V} \in \mathcal{O}_p \left( \mathbb{R} \right)$ and $\textbf{D} \in \mathcal{M}_{n,p} \left( \mathbb{R} \right)$ a rectangular matrix with singular values on the diagonal. Using left-multiplication of the initial model by $\textbf{W} \textbf{U}^T$ 
\begin{align*}
    \textbf{W} \textbf{U}^T \textbf{y} &= \textbf{W} \textbf{U}^T \textbf{Z} u + \textbf{W} \textbf{U}^T \textbf{e}.% \\
%    &= \textbf{W} \textbf{D} \textbf{V}^T u + \textbf{W} \textbf{U}^T \textbf{e}.
\end{align*}

\bigskip

Since $\textbf{U} \in \mathcal{O}_n (\mathbb{R})$ and $\textbf{W} \in \mathcal{O}_n (\mathbb{C})$ one can write
\begin{align*}
    \left\Vert \textbf{W} \textbf{U}^T \textbf{y} - \textbf{W} \textbf{U}^T \textbf{Z} u \right\Vert^2_2 + \lambda \left\Vert u \right\Vert^2_2 &= \left( \textbf{y} - \textbf{Z} u \right)^T \textbf{U} \textbf{W}^* \textbf{W} \textbf{U}^T \left( \textbf{y} - \textbf{Z} u \right) + \lambda \left\Vert u \right\Vert^2_2 \\
    &= \left\Vert \textbf{y} - \textbf{Z} u \right\Vert^2_2 + \lambda \left\Vert u \right\Vert^2_2     
\end{align*}
so $\lambda_{opt}$ is the same in the two models. 

\bigskip

The hat matrix in this new model is
\begin{align*}
    \tilde{\textbf{H}}_{\lambda} &= (\textbf{W} \textbf{U}^T \textbf{Z}) (\textbf{W} \textbf{U}^T \textbf{Z})^* \left( ( \textbf{W} \textbf{U}^T \textbf{Z}) (\textbf{W} \textbf{U}^T \textbf{Z})^* + \lambda \textbf{I}_n \right)^{-1} \\
    &= (\textbf{W} \textbf{D} \textbf{V}^T) (\textbf{W} \textbf{D} \textbf{V}^T)^* \left( ( \textbf{W} \textbf{D} \textbf{V}^T) (\textbf{W} \textbf{D} \textbf{V}^T)^* + \lambda \textbf{W} \textbf{W}^* \right)^{-1} \\
    &= \textbf{W} \textbf{D} \textbf{D}^T (\textbf{D}\textbf{D}^T + \lambda \textbf{I}_n )^{-1} \textbf{W}^* \in \text{Circ}(n)
\end{align*}

\bigskip

We showed that $\tilde{\textbf{H}}_{\lambda} \in \text{Circ}(n)$ and we approximate  $\forall i \in \llbracket 1,n \rrbracket,~ [h_{\lambda}]_{ii}$  by  $$\frac{1}{n} \text{tr} (\tilde{\textbf{H}}_{\lambda}) =\frac{1}{n} \overset{n}{\underset{k=1}{\sum}}  \frac{d_{k}^2}{d_{k}^2 + \lambda} =
\frac{1}{n} \text{tr} ({\textbf{H}}_{\lambda}) .  $$

\bigskip

Applying this in the expression of $\text{err}^{LOO}$ we have
\begin{align*}
    \text{err}^{GCV} &=\frac{\left\Vert \textbf{W} \textbf{U}^T \textbf{y} - \tilde{\textbf{H}}_{\lambda} \textbf{y} \right\Vert^2_2}{ \left[ \frac{1}{n} \text{tr} \left( \textbf{I}_{n} - \tilde{\textbf{H}}_{\lambda} \right) \right]^2 } \\
    &=\frac{\left\Vert \textbf{y} - \hat{\textbf{y}} \left( \lambda \right) \right\Vert^2_2}{ \left[ \frac{1}{n} \text{tr} \left( \textbf{I}_{n} - \textbf{H}_{\lambda} \right) \right]^2 }. %\\
%    &= \frac{1}{n} \textbf{y}^T ( \textbf{I}_n - \textbf{H}_{\lambda} ) ( \text{diag} (\textbf{I}_n - \textbf{H}_{\lambda}) )^{-2} ( \textbf{I}_n - \textbf{H}_{\lambda} ) \textbf{y}.
\end{align*} 

\subsection{Practical choice of $\lambda$}
\subsubsection{A grid of $\lambda$ for GCV}
An important issue in ridge regression is the search for the optimal $\lambda$.  A grid of $\lambda$ is often chosen empirically. In the context of the additive polygenic model, it is possible to use the link between heritability and $\lambda$ to determine a grid of $\lambda$:
\[ \underbrace{ \lbrace 0.01, 0.02, ..., 0.99 \rbrace }_{h^2_G} \rightarrow \underbrace{ \lbrace p \frac{1 - 0.01}{0.01}, p \frac{1 - 0.02}{0.02}, ..., p \frac{1 - 0.99}{0.99} \rbrace }_{\lambda}.\]

\subsubsection{Using Singular Value Decomposition to speed-up GCV computation}

Applying GCV with $n > p$, we would compute $\hat{u}_R = ( \textbf{Z}^T \textbf{Z} + \lambda \textbf{I}_p )^{-1} \textbf{Z}^T \textbf{y}$ using  for each $\lambda$ and use it to make prediction. In the context of GWAS (i.e. $p >> n$), this is not optimal since it implies the inversion of a $p \times p$ matrix. In our situation, the dual solution of ridge regression is much more adapted, leading to: 
\begin{align}
    \hat{\textbf{y}} \left( \lambda \right) = \textbf{Z} \hat{u}_{R} \left( \lambda \right) = \textbf{Z} \textbf{Z}^T \left( \textbf{Z} \textbf{Z}^T + \lambda \textbf{I}_n \right)^{-1} \textbf{y} = \textbf{H}_{\lambda} \textbf{y}.
\end{align}

\bigskip

GCV can be rewritten for more efficient computation. Let $\textbf{Z} = \textbf{U} \textbf{D} \textbf{V}^T$ be the singular value decomposition (SVD) of $\textbf{Z}$ with $\textbf{U} \in \mathcal{O}_n \left( \mathbb{R} \right)$, $\textbf{V} \in \mathcal{O}_p \left( \mathbb{R} \right)$ and $\textbf{D} \in \mathcal{M}_{n,p} \left( \mathbb{R} \right)$ a rectangular matrix with singular values on the diagonal, we have $\textbf{Z} \textbf{Z}^T = \textbf{U} \textbf{D} \textbf{D}^T \textbf{U}^T$ the eigen decomposition of $\textbf{Z} \textbf{Z}^T$. Rewriting $\textbf{H}_{\boldsymbol{\lambda}}$ using the SVD and applying it to GCV leads to:

\begin{align}
    \textbf{H}_{\boldsymbol{\lambda}} &= \textbf{U} \left[ \textbf{D} \textbf{D}^T \left( \textbf{D} \textbf{D}^T + \lambda \textbf{I}_n \right)^{-1} \right] \textbf{U}^T = \textbf{U} \boldsymbol{\mathcal{D}}_{\lambda} \textbf{U}^T, \\
    \text{err}^{GCV} &= \textbf{b}^T \left( \textbf{I}_n - \boldsymbol{\mathcal{D}}_{\lambda} \right) \left[ \frac{1}{n} \text{tr} \left( \textbf{I}_{\textbf{n}} - \boldsymbol{\mathcal{D}}_{\lambda} \right) \textbf{I}_{\textbf{n}} \right]^{-2} \left( \textbf{I}_n - \boldsymbol{\mathcal{D}}_{\lambda} \right) \textbf{b}, ~ \textbf{b} = \textbf{U}^T \textbf{y}.
\end{align}

\bigskip

Assuming that we have access to the eigen-decomposition of $\textbf{Z} \textbf{Z}^T$, the GCV computation as a function of diagonal matrices is extremely efficient. The most time-consuming part is the eigen decomposition (or the SVD).

\subsection{Issue with empirical scaling in the high dimensional case}
In this section we highlight the issue of the LOO / GCV with "naive" estimation of the intercept using empirically scaled matrices in the high dimensional context ( $n < p$ ). We first show why LOO does not work in this setup, then show why GCV does not work either and briefly highlight the issue for the "naive" estimation of more general fixed effects. 

\bigskip

In the following, we still consider the genotype matrix $\textbf{Z}$ and the phenotype vector \textbf{y} to be empirically scaled. One immediate consequence of this scaling is $\forall i \in \llbracket 1, n \rrbracket, y_i = - \sum_{j \neq i} y_j$ and $z_i = - \sum_{j \neq i} z_j$. Since each row of $\textbf{Z}$ is a linear combination of the others, we also have $0 \in \text{sp} \left( \textbf{ZZ}^T \right)$.

\subsubsection{Constant eigenvectors associated with the null eigenvalue\label{sec::vecteur_propre_constant}}
Let $\textbf{w} = \alpha \boldsymbol{\mathds{1}_n}$ with $\alpha \in \mathbb{R}$. Since $\textbf{Z}$ is normalized with the empirical scaling $\textbf{Z}^T \textbf{w} = \textit{0}_p \rightarrow \textbf{Z} \textbf{Z}^T \textbf{w} = \boldsymbol{0_n} = 0 \textbf{w}$ so the eigenvectors associated with 0 are constant.

\bigskip

Since we choose this eigenvector to have a  unit norm, we have $\left\Vert \textbf{w} \right\Vert^2_2 = \alpha^2 \left\Vert \boldsymbol{\mathds{1}_n} \right\Vert^2_2 = \alpha^2 n$. In the end, $\textbf{w} = \frac{1}{\sqrt{n}} \boldsymbol{\mathds{1}_n}$ or $\textbf{w} = \frac{-1}{\sqrt{n}} \boldsymbol{\mathds{1}_n}$.

\subsubsection{LOO standardization problem in a high dimensional setting\label{sec::pb_loo}}

Let $\textbf{Z}_{-i} = \begin{pmatrix}
z_1 \\
\vdots \\
z_{i-1} \\
z_{i+1} \\
\vdots\\
z_n
\end{pmatrix} \in \mathcal{M}_{n-1,p}\left(\mathbb{R}\right)$, $\textbf{y}_{-i} = \begin{pmatrix}
y_1 \\
\vdots \\
y_{i-1} \\
y_{i+1} \\
\vdots\\
y_{n}
\end{pmatrix}$  and $\hat{u}_R^{-i} = \textbf{Z}_{-i}^T \left( \textbf{Z}_{-i} \textbf{Z}_{-i}^T + \lambda \textbf{I}_{n-1} \right)^{-1} \textbf{y}_{-i} $. 

\bigskip

Then, we have $\hat{\textbf{y}}_{-i} (i) = z_i^T \hat{u}_R^{-i} = - \sum_{j \neq i} z_j^T \hat{u}_R^{-i} = - \mathds{1}_n^T \textbf{Z}_{-i} \textbf{Z}_{-i}^T \left( \textbf{Z}_{-i} \textbf{Z}_{-i}^T + \lambda \textbf{I}_{n-1} \right)^{-1} \textbf{y}_{-i}$.

\bigskip

We assume the variants to be independent and can reasonably suppose that the individuals are linearly independent  when $n<p$. In that case  $\textbf{Z}_{-i} \textbf{Z}_{-i}^T$ is invertible in spite of empirical centering because the empirical centering includes the i-th individual. We notice that when $\lambda \rightarrow 0$, $\hat{\textbf{y}}_{-i} (i) \rightarrow - \sum_{j \neq i} y_j = y_i$. Then, we easily show that $\left( \hat{\textbf{y}}_{-i} (i) - y_i \right)^2 \rightarrow 0$ and $\text{err}^{LOO} \underset{\lambda \rightarrow 0}{\rightarrow} 0$.

\bigskip

Here we see the influence from the unique standardisation of this LOO : because we used all individuals for standardization a phenomenon of dependency appears between the training and validation sets. Have we used a classical n-fold cross validation we would not have such dependencies, since the standardization would only include the training set.

\subsubsection{ GCV  standardization  problem  in  a  high  dimensional  setting\label{sec:limite_GCV}}
%$\text{lim}\limits_{\substack{d^2_n = 0 \\ \lambda %\rightarrow 0}} \text{GCV}$\label{sec:limite_GCV}}

Starting from GCV formula and assuming $n < p$

\begin{align*}
    \text{err}^{GCV} ( \textbf{y}, \textbf{Z}, \lambda ) &= \frac{1}{n} \textbf{y}^T ( \textbf{I}_n - \textbf{H}_{\lambda} ) \left[ \frac{1}{n} \text{tr}( \textbf{I}_n - \textbf{H}_{\lambda} ) \textbf{I}_n \right]^{-2} ( \textbf{I}_n - \textbf{H}_{\lambda} ) \textbf{y} \\
    &= \frac{1}{n} \textbf{b}^T ( \textbf{I}_n - \boldsymbol{\mathcal{D}}_{\lambda} ) \left[ \frac{1}{n} \text{tr}( \textbf{I}_n - \boldsymbol{\mathcal{D}}_{\lambda} ) \textbf{I}_n \right]^{-2} ( \textbf{I}_n - \boldsymbol{\mathcal{D}}_{\lambda} ) \textbf{b}
\end{align*}{}
where $\textbf{b} = \textbf{U}^T \textbf{y}$ and $\boldsymbol{\mathcal{D}}_{\lambda}=\textbf{D} \textbf{D}^T \left( \textbf{D} \textbf{D}^T + \lambda \textbf{I}_n \right)^{-1}.$

\bigskip

Let $d^2_n$ the null eigenvalue of $\textbf{ZZ}^T$ thus obtained. Noticing that $\textbf{I}_n - \boldsymbol{\mathcal{D}}_{\lambda} \xrightarrow[\lambda \rightarrow 0]{d^2_n = 0} \begin{pmatrix} 0 & & & \\ & \ddots & & \\ & & 0 & \\ & & & 1 \end{pmatrix}$, we have $\left[ \frac{1}{n} \text{tr}( \textbf{I}_n - \boldsymbol{\mathcal{D}}_{\lambda} ) \textbf{I}_n \right]^{-2} \xrightarrow[\lambda \rightarrow 0]{d^2_n = 0} n^2 \textbf{I}_n$. 

\bigskip

We then have 

\begin{align*}
    \text{err}^{GCV} ( \textbf{y}, \textbf{Z}, \lambda ) \xrightarrow[\lambda \rightarrow 0]{d^2_n = 0} \frac{1}{n} \times n^2 \textbf{b}^T \begin{pmatrix} 0 & & & \\ & \ddots & & \\ & & 0 & \\ & & & 1 \end{pmatrix} \textbf{b} = n b_n^2.
\end{align*}

\bigskip

Using \ref{sec::vecteur_propre_constant}, $b^2_n =  \frac{1}{n} ( \boldsymbol{\mathds{1}_n}^T \textbf{y} )^2$ and so 

\begin{align*}
    \text{err}^{GCV} ( \textbf{y}, \textbf{Z}, \lambda ) \xrightarrow[\lambda \rightarrow 0]{d^2_n = 0} ( \boldsymbol{\mathds{1}_n}^T \textbf{y} )^2 = 0.
\end{align*}

\bigskip

A similar issue can be observed in the presence of covariates. Let $\textbf{X} \in \mathcal{M}_{n,r}(\mathbb{R})$ the empirically scaled matrix of covariates. A "naive" approach to take into account those covariates would be to perform linear regression of the  phenotypes (which we assumed to be centered) on the empirically scaled covariates and then to apply GCV on the residuals. Let $\hat{\beta}$ the least square estimator, in this setup
\begin{align*}
    \text{err}^{GCV} ( \textbf{y} - \textbf{X} \hat{\beta} , \textbf{Z}, \lambda ) &\xrightarrow[\lambda \rightarrow 0]{d^2_n = 0} ( \boldsymbol{\mathds{1}_n}^T \textbf{y} -  \boldsymbol{\mathds{1}_n}^T \textbf{X} \hat{\beta} )^2 \\
    &= ( 0 - 0 )^2 \\
    &= 0.
\end{align*}

\subsection{Projection-based approach for GCV with covariates using QR decomposition }

QR decomposition allows an easy construction of a contrast matrix. The QR decomposition of $\textbf{A} \in \mathbb{M}_{n,r} ( \mathbb{R} ) $ is  $\textbf{A} = \textbf{Q} \textbf{R}$ with $\textbf{Q} \in \mathcal{O}(n)$ and $\textbf{R}^T \in \mathcal{M}_{r,n}( \mathbb{R} ) = [ \textbf{R}_1^T, \textbf{0}_{r,n-r}]$ where $\textbf{R}_1 \in \mathcal{M}_{r,r}(\mathbb{R})$ is an upper triangular matrix.

\bigskip

Let $\textbf{Q} = [\textbf{Q}_1, \textbf{Q}_2]$ with $\textbf{Q}_1 \in \mathcal{M}_{n,r}( \mathbb{R} ), ~\textbf{Q}_2 \in \mathcal{M}_{n,n-r}( \mathbb{R} )$ and observing that 
\begin{align*}
    \textbf{Q}^T \textbf{Q} = \begin{pmatrix} \textbf{Q}_1^T \\ \textbf{Q}_2^T \end{pmatrix} \begin{pmatrix} \textbf{Q}_1 & \textbf{Q}_2 \end{pmatrix} = \begin{pmatrix} \textbf{Q}_1^T \textbf{Q}_1 & \textbf{Q}_1^T \textbf{Q}_2 \\ \textbf{Q}_2^T \textbf{Q}_1 & \textbf{Q}_2^T \textbf{Q}_2 \end{pmatrix} = \begin{pmatrix} \textbf{I}_{r} & \textbf{O}_{r,n-r} \\ \textbf{O}_{n-r,r} & \textbf{I}_{n-r} \end{pmatrix},
\end{align*}{}
we can show 
\begin{align*}
    \textbf{Q}_2^T \textbf{A} =  \textbf{Q}_2^T \begin{pmatrix} \textbf{Q}_1 & \textbf{Q}_2 \end{pmatrix} \textbf{R} = \begin{pmatrix} \textbf{0}_{n-r,r} &\textbf{I}_{n-r} \end{pmatrix} \begin{pmatrix} \textbf{R}_1 \\ \textbf{0}_{n-r,n} \end{pmatrix} = \textbf{0}_{n-r,r}.
\end{align*}{}

\bigskip

$\textbf{Q}_2^T$ is a contrast matrix since we have $\textbf{Q}_2^T \textbf{A} =\textbf{0}_{n-r,r}$ and $\textbf{Q}_2^T \textbf{Q}_2 = \textbf{I}_{n-r}$. The QR decomposition of a matrix being relatively inexpensive to compute,  this proposed method offers an interesting alternative.

\subsection{Link between random effects model and ridge regression}

\subsubsection{The case without fixed effects}
Ridge regression and random effects model are closely linked. Starting by the maximizing the posterior of the parameters of 
\begin{align*}
     \textbf{y} = \textbf{Z} u + \textbf{e}
\end{align*}{}
where $u \sim \mathcal{N}(0_p, \tau \textbf{I}_p )$ and $\textbf{e} \sim \mathcal{N}(0_n, \sigma^2 \textbf{I}_n )$. Our goal is to maximize
\begin{align*}
    &p( u \vert \textbf{y} ) = \frac{ p( \textbf{y} \vert u ) p(u)}{p(\textbf{y})} \\
    \rightarrow &\text{log} ~ p( u \vert \textbf{y} )  = \text{log} ~  p( \textbf{y} \vert u ) + \text{log} ~  p(u) - \text{log} ~ p(\textbf{y}).
\end{align*}

\bigskip

Using the fact that $u \sim \mathcal{N}(0_p, \tau \textbf{I}_p )$, $\textbf{y} \sim \mathcal{N}(\textbf{0}_n, \tau \textbf{Z} \textbf{Z}^T + \sigma^2 \textbf{I}_n )$, $\textbf{y} \vert u \sim \mathcal{N}(\textbf{Z} u, \sigma^2 \textbf{I}_n )$ and remembering the formula of the log-likelihood for a gaussian distribution of parameters $\mu$ and $\boldsymbol{\Sigma}$ is
\begin{align*}
\text{log } p( \textbf{x} \vert \mu, \boldsymbol{\Sigma} ) = - \frac{n}{2} \text{log } 2 \pi - \frac{1}{2}  \text{log} \left\vert \boldsymbol{\Sigma} \right \vert - ( \textbf{x} - \mu )^T \boldsymbol{\Sigma}^{-1} ( \textbf{x} - \mu ),
\end{align*}
we can write
\begin{align*}
    \text{log} ~ p( u \vert \textbf{y} )  =& -\frac{n}{2} \text{log } 2 \pi - \frac{1}{2}  \text{log} \left\vert \sigma^2 \textbf{I}_n \right \vert - \frac{1}{2} ( \textbf{y} - \textbf{Z} u )^T ( \sigma^2 \textbf{I}_n )^{-1} ( \textbf{y} - \textbf{Z} u ) \\
    & -\frac{p}{2} \text{log } 2 \pi - \frac{1}{2}  \text{log} \left\vert \tau \textbf{I}_p \right \vert - \frac{1}{2} ( u - 0_p )^T (\tau \textbf{I}_p )^{-1} ( u - 0_p ) \\
    & + \frac{n}{2} \text{log } 2 \pi + \frac{1}{2}  \text{log} \left\vert \tau \textbf{Z} \textbf{Z}^T + \sigma^2 \textbf{I}_n \right \vert - \frac{1}{2} ( \textbf{y} - \textbf{0}_n )^T ( \tau \textbf{Z} \textbf{Z}^T + \sigma^2 \textbf{I}_n )^{-1} ( \textbf{y} - \textbf{0}_n ).
\end{align*}

\bigskip

By isolating the terms dependent on $u$, we obtain
\begin{align*}
    \text{log} ~ p( u \vert \textbf{y} )  =& - \frac{1}{2 \sigma^2} \left( \left\Vert \textbf{y} - \textbf{Z} u \right\Vert^2_2 + \frac{\sigma^2}{\tau} \left\Vert u \right\Vert^2_2 \right) + K_{\perp u}
\end{align*}
with $K_{\perp u}$ a term independent of $u$. After simplification we have
\begin{equation}
    \argmax_{u} p \left( u \vert \textbf{y} \right) = \argmin_{u} \left\Vert \textbf{y} - \textbf{Z} u \right\Vert^2_2 + \lambda \left\Vert u \right\Vert_2^2 \text{ with } \lambda = \frac{\sigma^2}{\tau}.
    \label{eq:equivalence_ridge_mm_sup}
\end{equation}

\subsubsection{An extension for the mixed model}
It is also possible to exhibit a link between mixed model (that is a random effects model with additional covariates with non-random effects) and ridge regression with some covariates we do not wish to penalize. Assuming the following model:
\begin{align*}
     \textbf{y} &=  \textbf{X} \beta +  \textbf{Z} u + \textbf{e}
\end{align*}
and denoting $\textbf{C}$ a contrast matrix such that $\textbf{C} \textbf{X} = \textbf{0}_{n,r}$ and $\textbf{C} \textbf{C}^T = \textbf{I}_{n-r}$.

\bigskip

The left multiplication of the above by $\textbf{C}$ gives
\begin{align*}
     \textbf{C} \textbf{y} &=  \textbf{C} \textbf{X} \beta +  \textbf{C} \textbf{Z} u +  \textbf{C} \textbf{e} =  \textbf{C} \textbf{Z} u +  \textbf{C} \textbf{e}. 
\end{align*}

\bigskip

Noticing that $ \textbf{C} \textbf{y} \vert u \sim \mathcal{N} (  \textbf{C} \textbf{Z} u, \sigma^2 \textbf{I}_{n-r} )$ and $ \textbf{C} \textbf{y} \sim \mathcal{N} (  \textbf{0}_{n-r}, \tau \textbf{C} \textbf{Z} \textbf{Z}^T \textbf{C}^T + \sigma^2 \textbf{I}_{n-r} )$ we can write the posterior of the contrasted model as
\begin{align*}
    \text{log} ~ p( u \vert \textbf{C} \textbf{y} )  =& - \frac{1}{2} \left( \left\Vert \textbf{C} \textbf{y} - \textbf{C} \textbf{Z} u \right\Vert^2_2 + \frac{\sigma}{\tau} \left\Vert u \right\Vert^2_2 \right) + K_{\perp u}
\end{align*}
and after simplification
\begin{equation*}
    \argmax_{u} p \left( u \vert \textbf{C} \textbf{y} \right) = \argmin_{u} \left\Vert \textbf{C} \textbf{y} - \textbf{C} \textbf{Z} u \right\Vert^2_2 + \lambda \left\Vert u \right\Vert_2^2 \text{ with } \lambda = \frac{\sigma^2}{\tau}.
\end{equation*}

\subsection{The proportion of causal variants does not impact heritability estimation}

\subsubsection{Estimation of heritability on synthetic data}
\begin{figure}[!h]
    \centering
    \includegraphics[width=0.7\linewidth]{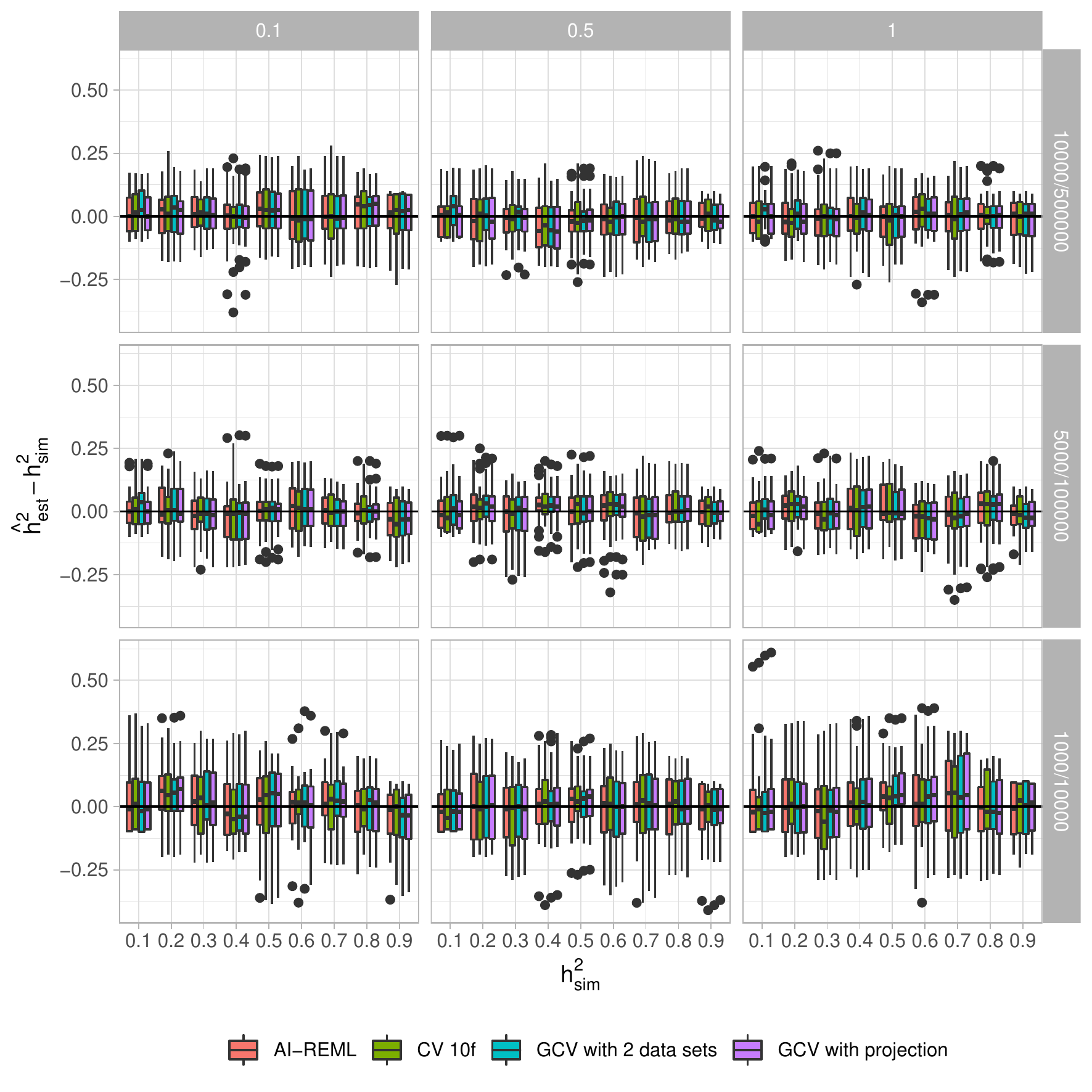}
    \caption{Distribution of $(h^2_{est} - h^2_{sim})$  for multiple parameter combinations with 30 replications. Data are simulated under the fully synthetic procedure. The columns of the grid correspond to the fraction of causal variants while the lines correspond to the ratio $n/p$. For each panel the horizontal axis corresponds to the simulated heritability $h^2_{sim} \in \lbrace 0.1, ...,\rbrace$ and the vertical axis corresponds to the estimation of $h^2_g - h^2_{sim}$. Heritability estimations are done with random effects model using AI-REML to estimate the variance components and with ridge regression using 3 approaches for the choice of $\lambda_{opt}$ : GCV with a projection correction and GCV with a 2nd dataset correction and a 10 fold cross validation ( Ridge 10fCV ).}
    \label{grid_boxplot_ecart_h2g}
\end{figure}{}

\newpage

\subsubsection{Estimation of heritability on semi-synthetic data}
\begin{figure}[!h]
    \centering
    \includegraphics[width=0.7\linewidth]{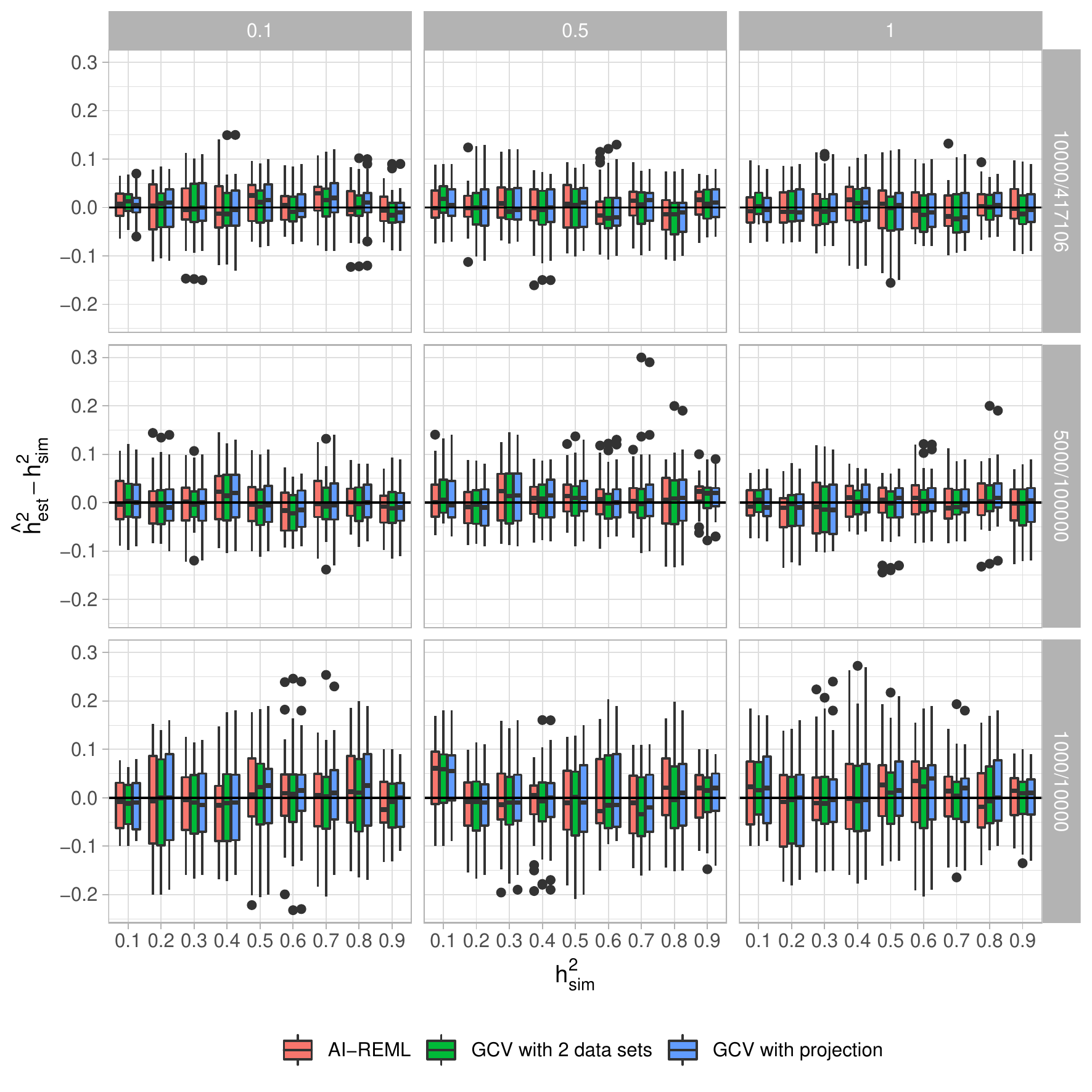}
    \caption{Distribution of $(h^2_{est} - h^2_{sim})$  for multiple parameter combinations with 30 replications. Data are simulated under the semi-synthetic procedure. The columns of the grid correspond to the fraction of causal variants while the lines correspond to the ratio $n/p$. For each panel the horizontal axis corresponds to the simulated heritability $h^2_{sim} \in \lbrace 0.1, ...,\rbrace$ and the vertical axis corresponds to the estimation of $h^2_g - h^2_{sim}$. Heritability estimations are done with random effects model using AI-REML to estimate the variance components and with ridge regression using 2 approaches for the choice of $\lambda_{opt}$ : GCV with a projection correction and GCV with a 2nd dataset correction.}
    \label{grid_boxplot_avec_LD_h2g}
\end{figure}

\newpage

\subsection{Approximation of predictive power}

In this section we detail our approximation of the MSE and squared correlation. In the following the index $_ {tr}$ refers to the training set whereas $_ {te}$ refers to the test set. To lighten notations $\textbf{Z}$ is the normalized genotype matrix of the training set. Let $z_{te} \in \mathbb{R}^p$ the column vector corresponding to the normalized genotypes of one test individual. We assume $\mathbb{E}_{z_{te}} \left[ z_{te} \right] = 0_p$ and $\text{var}(z_{te}) = \textbf{I}_p$.

\bigskip

We remind that $\hat{u} = \textbf{Z}^T \left( \textbf{Z} \textbf{Z}^T + \lambda \textbf{I}_n \right)^{-1} \textbf{y}_{tr} = \left( \textbf{Z}^T \textbf{Z}  + \lambda \textbf{I}_p \right)^{-1} \textbf{Z}^T \textbf{y}_{tr} = \textbf{K}_{\lambda} \textbf{y}_{tr}$. We assume the phenotype to have unit variance without loss of generality.  
\bigskip

Lastly we remind that for $\textbf{x}$ a random vector with $\mathbb{E}[\textbf{x}] = \textbf{a}$ and $\text{var}(\textbf{x}) = \textbf{V}$ we have for any matrix $\textbf{B}$ $\mathbb{E}[\textbf{x}^T \textbf{B} \textbf{x}] = \text{tr} ( \textbf{B} \textbf{V} ) + \textbf{a}^T \textbf{B} \textbf{a}$.

\bigskip 

\subsubsection{Approximation of the mean squared error on the test set}

Using the classic bias-variance decomposition and assuming $\textbf{Z}$ fixed, we can write
\begin{align*}
    \mathbb{E}_{\textbf{y}_{tr}, y_{te}, z_{te}} \left[ ( y_{te} - \hat{y}_{te} )^2 \right] &= \mathbb{E}_{z_{te}} \left[ \mathbb{E}_{\textbf{y}_{tr}, y_{te} \vert z_{te}} \left[ ( y_{te} - \hat{y}_{te} )^2 \right] \right] \\
    &= \mathbb{E}_{z_{te}} \left[ \text{var}(y_{te} \vert z_{te}) + \text{var}(\hat{y}_{te} \vert z_{te} ) + \left( \mathbb{E}_{\textbf{y}_{tr} \vert z_{te}} \left[ \hat{y}_{te} \right] - \mathbb{E}_{y_{te} \vert z_{te}} [ y_{te} ] \right)^2 \right].
\end{align*}

\bigskip

Firstly we have
\begin{align*}
    \mathbb{E}_{ y_{te}, z_{te} } [ y_{te} ] &=  \mathbb{E}_{ y_{te}, z_{te} } [ z_{te}^T u + e_{te} ] = \mathbb{E}_{z_{te}^T} [ z_{te} ] u + \mathbb{E}_{e_{te}} [ e_{te} ] = 0. \\
    \mathbb{E}_{ \textbf{y}_{tr} } [ y_{te} ] &=  y_{te}. \\
    \mathbb{E}_{ y_{te}, z_{te} } [ \hat{y}_{te} ] &=  \mathbb{E}_{ y_{te}, z_{te} } [ z_{te}^T \textbf{K}_{\lambda} \textbf{y}_{tr} ] = 0. \\
    \mathbb{E}_{ \textbf{y}_{tr} } [ \hat{y}_{te} ] &=  \mathbb{E}_{ \textbf{y}_{tr} } [ z_{te}^T \textbf{K}_{\lambda} \textbf{y}_{tr} ] = z_{te}^T \textbf{K}_{\lambda} \textbf{Z} u.
\end{align*}

\bigskip

Developing the 3 terms of the bias-variance decomposition using the expected value of a quadratic form we have 
\begin{align*}
    \text{var}(y_{te} \vert z_{te}) &= \mathbb{E}_{y_{te} \vert z_{te}} \left[ \left( y_{te} -  \mathbb{E}_{y_{te} \vert z_{te}} \left[ y_{te} \right] \right)^2\right] = \mathbb{E}_{y_{te} \vert z_{te}} \left[ e_{te}^2 \right] = \sigma^2 \\
    \text{var}(\hat{y}_{te} \vert z_{te} ) &=  \mathbb{E}_{\textbf{y}_{tr} \vert z_{te}} \left[ \hat{y}_{te}^2 \right] - \mathbb{E}_{\textbf{y}_{tr} \vert z_{te}} \left[ \hat{y}_{te} \right]^2 \\
    &= \mathbb{E}_{\textbf{y}_{tr} \vert z_{te}} \left[  \textbf{y}_{tr}^T \textbf{K}_{\lambda}^T z_{te} z_{te}^T \textbf{K}_{\lambda} \textbf{y}_{tr} \right] -  ( \textbf{Z} u )^T \textbf{K}_{\lambda}^T z_{te} z_{te}^T \textbf{K}_{\lambda} ( \textbf{Z} u )  \\
    &= \text{tr} \left( \textbf{K}_{\lambda}^T z_{te} z_{te}^T \textbf{K}_{\lambda} \sigma^2 \textbf{I}_n \right) + ( \textbf{Z} u )^T \textbf{K}_{\lambda}^T z_{te} z_{te}^T \textbf{K}_{\lambda} ( \textbf{Z} u ) \\
    & ~- ( \textbf{Z} u )^T \textbf{K}_{\lambda}^T z_{te} z_{te}^T \textbf{K}_{\lambda} ( \textbf{Z} u ) \\
    &= \sigma^2 z_{te}^T \textbf{K}_{\lambda} \textbf{K}_{\lambda}^T z_{te} \\
    \left( \mathbb{E}_{\textbf{y}_{tr} \vert z_{te}} \left[ \hat{y}_{te} \right] - \mathbb{E}_{y_{te} \vert z_{te}} [ y_{te} ] \right)^2 &= \left( z_{te}^T \textbf{K}_{\lambda} ( \textbf{Z} u ) - z_{te}^T u \right)^2 \\
    &= \left( z_{te}^T \left( \textbf{K}_{\lambda} \textbf{Z} - \textbf{I}_{p} \right) u \right)^2 \\
    &= z_{te}^T \left( \textbf{K}_{\lambda} \textbf{Z} - \textbf{I}_{p} \right) u u^T \left( \textbf{K}_{\lambda} \textbf{Z} - \textbf{I}_{p} \right) z_{te}.
\end{align*}
since $ ( \textbf{K}_{\lambda} \textbf{Z} )^T = ( \textbf{Z}^T ( \textbf{Z} \textbf{Z}^T + \lambda \textbf{I}_n )^{-1} \textbf{Z} )^T = \textbf{Z}^T ( \textbf{Z} \textbf{Z}^T + \lambda \textbf{I}_n )^{-1} \textbf{Z} = \textbf{K}_{\lambda} \textbf{Z}$.

\bigskip

Applying the expectation over $z_{te}$ on those 3 terms
\begin{align*}
    \mathbb{E}_{z_{te}} \left[ \text{var}(y_{te} \vert z_{te}) \right] &= \sigma^2 \\
    \mathbb{E}_{z_{te}} \left[ \text{var}(\hat{y}_{te} \vert z_{te} ) \right] &= \sigma^2 \text{tr} \left( \textbf{K}_{\lambda} \textbf{K}_{\lambda}^T \right) \\
    \mathbb{E}_{z_{te}} \left[ \left( \mathbb{E}_{\textbf{y}_{tr} \vert z_{te}} \left[ \hat{y}_{te} \right] - \mathbb{E}_{y_{te} \vert z_{te}} [ y_{te} ] \right)^2 \right] &= u^T \left( \textbf{K}_{\lambda} \textbf{Z} - \textbf{I}_{p} \right)^2 u \\
    &= u^T \left( \textbf{K}_{\lambda} \textbf{Z} \textbf{K}_{\lambda} \textbf{Z} - 2 \textbf{K}_{\lambda} \textbf{Z} + \textbf{I}_{p} \right) u.
\end{align*}

\bigskip

We first approximate the case $n < p$. Here we can reasonably suppose that $\textbf{Z} \textbf{Z}^T \simeq p \textbf{I}_n$ since we are working on unrelated individuals and because of the normalization of $\textbf{Z}$. Using this approximation one can write

\begin{align*}
    \textbf{Z} \textbf{Z}^T \simeq p \textbf{I}_n &\Rightarrow \textbf{K}_{\lambda} \simeq \frac{1}{p + \lambda} \textbf{Z}^T.
\end{align*}

\bigskip

Replacing $\textbf{Z} \textbf{Z}^T$ in the above expressions and using the link between ridge regression parameter and heritability we have

\begin{align*}
    \mathbb{E}_{z_{te}} \left[ \text{var}(\hat{y}_{te} \vert z_{te} ) \right] &= \sigma^2 \text{tr} \left( \textbf{K}_{\lambda} \textbf{K}_{\lambda}^T \right) \\ 
    &\simeq \sigma^2 \text{tr} \left(  \left( \frac{1}{p + \lambda} \textbf{Z}^T \right) \left( \frac{1}{p + \lambda} \textbf{Z}^T \right)^T \right) \\
    &= \sigma^2 \left( \frac{1}{p + \lambda} \right)^2 \text{tr} \left(  \textbf{Z}^T \textbf{Z}  \right) \\
    &\simeq \sigma^2 \left( \frac{1}{p + \lambda} \right)^2 \text{tr} \left( p \textbf{I}_n \right) \\
    &=  \sigma^2 \left( \frac{1}{p + \lambda} \right)^2 n p \\
    &= (1 - h^2) (h^2)^2 \frac{n}{p}
\end{align*}
and
\begin{align*}
    \mathbb{E}_{z_{te}} \left[ \left( \mathbb{E}_{\textbf{y}_{tr} \vert z_{te}} \left[ \hat{y}_{te} \right] - \mathbb{E}_{y_{te} \vert z_{te}} [ y_{te} ] \right)^2 \right] &= u^T \left( \textbf{K}_{\lambda} \textbf{Z} - \textbf{I}_{p} \right)^2 u \\
    &= u^T \left( \left( \frac{1}{p + \lambda} \right)^2 \textbf{Z}^T \textbf{Z} \textbf{Z}^T \textbf{Z} - 2 \left( \frac{1}{p + \lambda} \right) \textbf{Z}^T \textbf{Z} + \textbf{I}_p \right) u \\
    &= p \left( \frac{1}{p + \lambda} \right)^2 ( \textbf{Z} u )^T ( \textbf{Z} u ) - 2 \left( \frac{1}{p + \lambda} \right) ( \textbf{Z} u )^T ( \textbf{Z} u ) + u^T u \\
    &\simeq p \left( \frac{1}{p + \lambda} \right)^2 n h^2  - 2 \left( \frac{1}{p + \lambda} \right) n h^2 + h^2 \\
    &= \frac{n}{p} (h^2)^3 - 2 \frac{n}{p} (h^2)^2 + h^2 \\
    &= h^2 \left( 1 + \frac{n}{p} \left( (h^2)^2 - 2 h^2 \right) \right).
\end{align*}

\bigskip

Summing all those expressions, we end up with

\begin{align*}
    \mathbb{E}_{\textbf{y}_{tr}, y_{te}, z_{te}} \left[ ( y_{te} - \hat{y}_{te} )^2 \right] &\simeq 1 - h^2 + (1 - h^2) (h^2)^2 \frac{n}{p} + h^2 \left( 1 + \frac{n}{p} \left( (h^2)^2 - 2 h^2 \right) \right) \\
    &= 1 + h^2 \left( -1 + (1 - h^2) h^2 \frac{n}{p} + 1 + \frac{n}{p} \left( (h^2)^2 - 2 h^2 \right) \right) \\
    &= 1 - \frac{n}{p} (h^2)^2.
\end{align*}

\bigskip

We now consider the case $n > p$. Here one the other hand we can reasonably suppose that $\textbf{Z}^T \textbf{Z} \simeq n \textbf{I}_p$ since we assume the genotypes to be independent and again because of the normalization of $\textbf{Z}$. Using this approximation one can write

\begin{align*}
    \textbf{Z}^T \textbf{Z}  \simeq n \textbf{I}_p &\Rightarrow \textbf{K}_{\lambda} \simeq \frac{1}{n + \lambda} \textbf{Z}^T  \\
\end{align*}

\bigskip

First noticing the following algebra

\begin{align*}
    \frac{n}{n+\lambda} = \frac{n}{n + p \frac{1-h^2}{h^2}} = \frac{\frac{n}{p}}{\frac{n}{p} + \frac{1-h^2}{h^2}} = \frac{\frac{n}{p} \times h^2}{n/p \times h^2 + (1-h^2)} = \frac{\frac{n}{p} \times h^2}{1 + h^2 \times ( \frac{n}{p} - 1 ) }
\end{align*}
\begin{align*}
    \frac{\lambda}{n+\lambda} = \frac{p \frac{1-h^2}{h^2}}{n + p \frac{1-h^2}{h^2}} = \frac{1-h^2}{\frac{n}{p} \times h^2 + (1-h^2)} = \frac{1-h^2}{1 + h^2 ( \frac{n}{p} -1 )}
\end{align*}
and replacing $\textbf{Z}^T \textbf{Z}$ by $n \textbf{I}_p$ we now have
\begin{align*}
    \mathbb{E}_{z_{te}} \left[ \text{var}(\hat{y}_{te} \vert z_{te} ) \right] &= \sigma^2 \text{tr} \left( \textbf{K}_{\lambda} \textbf{K}_{\lambda}^T \right) \\ 
    &\simeq \sigma^2 \text{tr} \left( \left( \frac{1}{n+\lambda} \right)^2 \textbf{Z}^T \textbf{Z} \right) \\ 
    &\simeq \sigma^2 \left( \frac{1}{n+\lambda} \right)^2 n p \\ 
    &=\sigma^2 \frac{1}{\frac{n}{p}} \left( \frac{\frac{n}{p} \times h^2}{1 + h^2 \times ( \frac{n}{p} - 1 ) } \right)^2
\end{align*}
\begin{align*}
    \mathbb{E}_{z_{te}} \left[ \left( \mathbb{E}_{\textbf{y}_{tr} \vert z_{te}} \left[ \hat{y}_{te} \right] - \mathbb{E}_{y_{te} \vert z_{te}} [ y_{te} ] \right)^2 \right]
    &= u^T \left( \textbf{K}_{\lambda} \textbf{Z} \textbf{K}_{\lambda} \textbf{Z} - 2 \textbf{K}_{\lambda} \textbf{Z} + \textbf{I}_{p} \right) u \\
    &\simeq u^T \left( \left( \frac{1}{n+\lambda} \right)^2 \textbf{Z}^T \textbf{Z} \textbf{Z}^T \textbf{Z} - 2 \left( \frac{1}{n+\lambda} \right) \textbf{Z}^T \textbf{Z} + \textbf{I}_{p} \right) u \\
    &\simeq \left( \frac{n}{n+\lambda} - 1 \right)^2 u^T  u \simeq \left( \frac{n}{n+\lambda} - 1 \right)^2 h^2 \\
    &= \left( \frac{1-h^2}{1 + h^2 ( \frac{n}{p} -1 )} \right)^2 h^2 
\end{align*}

\bigskip

Summing all those expressions, we end up with

\begin{align*}
    \mathbb{E}_{\textbf{y}_{tr}, y_{te}, z_{te}} \left[ ( y_{te} - \hat{y}_{te} )^2 \right] &\simeq (1 - h^2) \frac{1 + \frac{n}{p} h^2}{1 + h^2(\frac{n}{p} - 1)}
\end{align*}

\bigskip

In the end we have

\begin{align}
    \mathbb{E}_{\textbf{y}_{tr}, y_{te}, z_{te}} \left[ ( y_{te} - \hat{y}_{te} )^2 \right] \simeq \left\{ \begin{array}{l}
        1 - \frac{n}{p} ( h^2 )^2 \text{ si } n<p \\ 
        (1 - h^2) \frac{1 + \frac{n}{p} h^2}{1 + h^2(\frac{n}{p} - 1)} \text{ otherwise.}
        \end{array} \right.
\end{align}

\subsubsection{Approximation of the mean squared error on the training set}

We quickly remind our approximations

\begin{align*}
    \textbf{H}_{\lambda} = \textbf{Z} \textbf{K}_{\lambda} \simeq \left\{ \begin{array}{l}
    \frac{1}{p + \lambda} \textbf{Z} \textbf{Z}^T \simeq \frac{p}{p + \lambda} \textbf{I}_n \simeq h^2 \textbf{I}_n \text{ if } n<p  \\
    \frac{1}{n + \lambda} \textbf{Z} \textbf{Z}^T \text{ otherwise. }
    \end{array} \right.
\end{align*}

\bigskip

Assuming $\textbf{Z}$ is fixed and writing the expectation over $\textbf{y}_{tr}$ of the mean squared error on the training set, we have

\begin{align*}
    \mathbb{E}_{\textbf{y}_{tr}} \left[ \frac{1}{n} ( \textbf{y}_{tr} - \hat{\textbf{y}}_{tr} )^T ( \textbf{y}_{tr} - \hat{\textbf{y}}_{tr} ) \right] &= \mathbb{E}_{\textbf{y}_{tr}} \left[ \frac{1}{n} \textbf{y}_{tr}^T ( \textbf{I}_n - \textbf{H}_{\lambda} )^2 \textbf{y}_{tr} \right] \\
    &= \frac{1}{n} \left( \text{tr} ( ( \textbf{I}_n - \textbf{H}_{\lambda} )^2 \times \sigma^2 \textbf{I}_n ) + ( \textbf{Z} u )^T ( \textbf{I}_n - \textbf{H}_{\lambda} )^2 ( \textbf{Z} u ) \right).
\end{align*}
    
\textbf{}    

The focus is the approximation of $( \textbf{I}_n - \textbf{H}_{\lambda} )^2$ :   
    
\begin{align*}
    ( \textbf{I}_n - \textbf{H}_{\lambda} )^2 \simeq \left\{ \begin{array}{l}
        ( 1 - h^2 )^2\textbf{I}_n \text{ if } n<p  \\
        \textbf{I}_n - 2 \times  \frac{1}{n + \lambda} \textbf{Z} \textbf{Z}^T + \left( \frac{1}{n + \lambda} \right)^2 \textbf{Z} \textbf{Z}^T \textbf{Z} \textbf{Z}^T \simeq \textbf{I}_n - \frac{2}{n + \lambda} \textbf{Z} \textbf{Z}^T + \frac{n}{(n + \lambda)^2} \textbf{Z} \textbf{Z}^T \text{ else.}
    \end{array}\right.
\end{align*}

\textbf{}

The approximation is  straightforward for $n<p$. We thus focus on the  $n>p$ case. 

\begin{align*}
    \text{tr} \left( \textbf{I}_n - \frac{2}{n + \lambda} \textbf{Z} \textbf{Z}^T + \frac{n}{(n + \lambda)^2} \textbf{Z} \textbf{Z}^T \right) &= n - \frac{2}{n + \lambda} \text{tr} (\textbf{Z} \textbf{Z}^T) + \frac{n}{(n + \lambda)^2} \text{tr} (\textbf{Z} \textbf{Z}^T) \\
    &\simeq n - 2 p \frac{n}{n + \lambda} + p \left( \frac{n}{n + \lambda} \right)^2
\end{align*}
\begin{align*}
    &( \textbf{Z} u )^T ( \textbf{Z} u ) \simeq n h^2 \\
    &( \textbf{Z} u )^T ( - \frac{2}{n + \lambda} \textbf{Z} \textbf{Z}^T )( \textbf{Z} u ) =  - \frac{2}{n + \lambda} u^T \textbf{Z}^T \textbf{Z} \textbf{Z}^T \textbf{Z} u \simeq - 2 n \frac{n}{n + \lambda} u^T u \simeq - 2 n \frac{n}{n + \lambda} h^2 \\
    &( \textbf{Z} u )^T ( \frac{n}{(n + \lambda)^2} \textbf{Z} \textbf{Z}^T )( \textbf{Z} u ) = \frac{n}{(n + \lambda)^2} u^T \textbf{Z}^T \textbf{Z} \textbf{Z}^T \textbf{Z} u \simeq n \left( \frac{n}{n + \lambda} \right)^2 h^2.
\end{align*}

\bigskip

Factorizing those results according to $\frac{n}{n + \lambda}$ and $\left( \frac{n}{n + \lambda} \right)^2$ and using the algebras described above we end up with
\begin{align*}
    \mathbb{E}_{\textbf{y}_{tr}} \left[ \frac{1}{n} ( \textbf{y}_{tr} - \hat{\textbf{y}}_{tr} )^T ( \textbf{y}_{tr} - \hat{\textbf{y}}_{tr} ) \right] &\simeq \left\{ \begin{array}{l}
        ( 1 - h^2 )^2 \text{ if } n<p  \\
        1 - 2 \frac{n}{n + \lambda} \left( \frac{p}{n} ( 1 - h^2 ) + h^2 \right) + \left( \frac{n}{n + \lambda} \right)^2 \left( \frac{p}{n} ( 1 - h^2 ) + h^2 \right) \text{ otherwise.}
    \end{array}\right.
\end{align*}

\bigskip

\begin{figure}
  \centering
\includegraphics[width=0.7\linewidth]{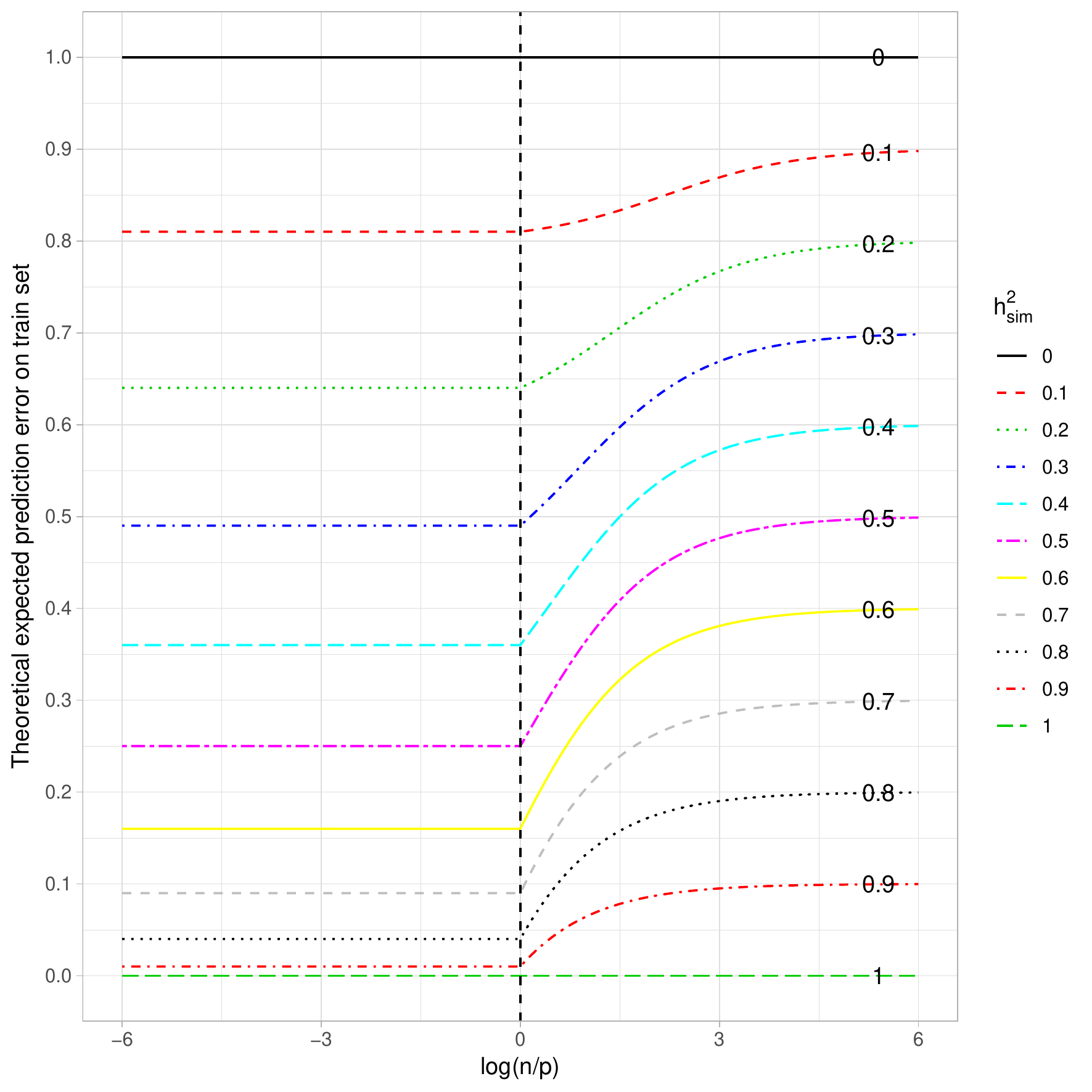}
  \caption{\label{fig:MSE-train-theo} Theoretical quadratic error on the training set with respect to  the log ratio of the number of individuals over the number of variants in the training set. Each curve corresponds to a given heritability (in the narrow sense). Note that the total variance is assumed to be 1.}
\end{figure}

\subsubsection{Approximation of the squared correlation on the test set}

Here we will explain our approximation of the correlation between the phenotype and the prediction. Assuming $\textbf{Z}$ is fixed, the correlation is
\begin{align*}
    \text{corr}( \hat{y}_{te}, y_{te}) = \frac{\text{cov}_{\textbf{y}_{tr}, y_{te}, z_{te}} ( y_{te}, \hat{y}_{te} )}{ \sqrt{\text{var}_{\textbf{y}_{tr}, y_{te}, z_{te}} \left[ y_{te} \right]} \sqrt{\text{var}_{\textbf{y}_{tr}, y_{te}, z_{te}} \left[ \hat{y}_{te} \right]} }.
\end{align*}

\bigskip

Estimating each of those 3 terms:
\begin{align*}
    \text{var}_{\textbf{y}_{tr}, y_{te}, z_{te}} \left[ \hat{y}_{te} \right] &= \mathbb{E}_{\textbf{y}_{tr}, y_{te}, z_{te}} [ \hat{y}_{te}^2 ] - \mathbb{E}_{\textbf{y}_{tr}, y_{te}, z_{te}} [ \hat{y}_{te} ]^2 \\
    &= \mathbb{E}_{\textbf{y}_{tr}, y_{te}, z_{te}} [ z_{te}^T \textbf{K}_{\lambda} \textbf{y}_{tr} \textbf{y}_{tr}^T \textbf{K}_{\lambda}^T z_{te}  ] - \mathbb{E}_{\textbf{y}_{tr}, y_{te}, z_{te}} [ z_{te}^T \textbf{K}_{\lambda} \textbf{y}_{tr} ]^2 \\
    &= \mathbb{E}_{\textbf{y}_{tr}} \mathbb{E}_{y_{te}, z_{te} \mid \textbf{y}_{tr}} [ z_{te}^T \textbf{K}_{\lambda} \textbf{y}_{tr} \textbf{y}_{tr}^T \textbf{K}_{\lambda}^T z_{te} ] - \mathbb{E}_{\textbf{y}_{tr}} \mathbb{E}_{y_{te}, z_{te} \mid \textbf{y}_{tr}} [ z_{te}^T \textbf{K}_{\lambda} \textbf{y}_{tr} ]^2 \\
    &= \mathbb{E}_{\textbf{y}_{tr}} [ \text{tr} ( \textbf{K}_{\lambda} \textbf{y}_{tr} \textbf{y}_{tr}^T \textbf{K}_{\lambda}^T ) + 0  ] - 0 \\
    &= \mathbb{E}_{\textbf{y}_{tr}} [ \textbf{y}_{tr}^T \textbf{K}_{\lambda}^T \textbf{K}_{\lambda} \textbf{y}_{tr} ] \\
    &= \text{tr}( \textbf{K}_{\lambda}^T \textbf{K}_{\lambda} \times \sigma^2 \textbf{I}_n ) + ( \textbf{Z} u )^T \textbf{K}_{\lambda}^T \textbf{K}_{\lambda} ( \textbf{Z} u )
\end{align*}

\begin{align*}
    \text{cov}_{\textbf{y}_{tr}, y_{te}, z_{te}} ( y_{te}, \hat{y}_{te} ) &= \mathbb{E}_{\textbf{y}_{tr}, y_{te}, z_{te}} [ ( y_{te} - \mathbb{E}_{\textbf{y}_{tr}, y_{te}, z_{te}} [ y_{te} ] ) ( \hat{y}_{te} - \mathbb{E}_{\textbf{y}_{tr}, y_{te}, z_{te}} [ \hat{y}_{te} ] ) ] \\
    &= \mathbb{E}_{\textbf{y}_{tr}, y_{te}, z_{te}} [ ( y_{te} - \mathbb{E}_{\textbf{y}_{tr}, y_{te}, z_{te}} [ z_{te}^T u + e_{te} ] ) ( \hat{y}_{te} - \mathbb{E}_{\textbf{y}_{tr}, y_{te}, z_{te}} [ z_{te}^T \textbf{K}_{\lambda} \textbf{y}_{tr} ] ) ] \\ 
    &= \mathbb{E}_{\textbf{y}_{tr}, y_{te}, z_{te}} [ ( y_{te} \hat{y}_{te} ) ]  \\
    &= \mathbb{E}_{y_{te}, z_{te}} \mathbb{E}_{\textbf{y}_{tr} \mid y_{te}, z_{te}} [ y_{te} z_{te}^T \textbf{K}_{\lambda} \textbf{y}_{tr} ] \\
    &= \mathbb{E}_{y_{te}, z_{te}} [ y_{te} z_{te}^T \textbf{K}_{\lambda} \textbf{Z} u ] \\
    &= \mathbb{E}_{y_{te}, z_{te}} [ z_{te}^T \textbf{K}_{\lambda} \textbf{Z} u ( u^T z_{te} + e_{te}^T ) ] \\
    &= \mathbb{E}_{y_{te}, z_{te}} [ z_{te}^T \textbf{K}_{\lambda} \textbf{Z} u u^T z_{te} ] + \mathbb{E}_{y_{te}, z_{te}} [ z_{te}^T \textbf{K}_{\lambda} \textbf{Z} u \times e_{te}^T ] \\ 
    &= \text{tr} ( \textbf{K}_{\lambda} \textbf{Z} u u^T ) + 0 + 0 ~ \left( z_{te}^T \perp e_{te},~ \mathbb{E} [e_{te}] = 0 ~\right) \\
    &= u^T \textbf{K}_{\lambda} \textbf{Z} u
\end{align*}

\begin{align*}
    \text{var}_{\textbf{y}_{tr}, y_{te}, z_{te}} \left[ y_{te} \right] &= \mathbb{E}_{\textbf{y}_{tr}, y_{te}, z_{te}} [ y_{te}^2 ] - \mathbb{E}_{\textbf{y}_{tr}, y_{te}, z_{te}} [ y_{te} ]^2 \\
    &= \mathbb{E}_{y_{te}, z_{te}} [ ( z^T_{te} u + e_{te} )^2 ] - 0 \\
    &= \mathbb{E}_{e_{te}, z_{te}}  \left[ ( z_{te}^T u )^2 \right] + \mathbb{E}_{e_{te}, z_{te}} \left[ (e_{te} )^2 \right] + 2 \mathbb{E}_{e_{te}, z_{te}} \left[ (z_{te}^T u) e_{te} \right] \\
    &= \mathbb{E}_{z_{te}} [ z_{te}^T u u^T z_{te} ]+ \mathbb{E}_{e_{te}} [ e_{te}^2 ]+ 2 \mathbb{E}_{z_{te}} [ z_{te}^T u ] \mathbb{E}_{e_{te}} [ e_{te} ] \\
    &= u^T u + \sigma^2 + 0 \\
\end{align*}

\bigskip

We  replace the empirical covariance matrices by their respective approximation according to the cases $n<p$ and $n>p$.

\bigskip

The $n<p$ case :
\begin{align*} 
    \textbf{Z} \textbf{Z}^T \simeq p \textbf{I}_n &\Rightarrow \textbf{K}_{\lambda} \simeq \frac{1}{p + \lambda} \textbf{Z}^T \Rightarrow  \textbf{K}_{\lambda}^T \textbf{K}_{\lambda} \simeq \frac{(h^2)^2}{p} \textbf{I}_n
\end{align*}

\begin{align*}
    &\sigma^2 \times \text{tr}( \textbf{K}_{\lambda}^T \textbf{K}_{\lambda} ) \simeq \frac{n}{p} (h^2)^2 ( 1 - h^2 ) \\
    &( \textbf{Z} u )^T \textbf{K}_{\lambda}^T \textbf{K}_{\lambda} ( \textbf{Z} u ) \simeq \frac{(h^2)^2}{p} ( \textbf{Z} u )^T ( \textbf{Z} u ) \simeq \frac{n}{p} (h^2)^2 \times h^2 \\
    &u^T \textbf{K}_{\lambda} \textbf{Z} u \simeq u^T  \frac{1}{p + \lambda} \textbf{Z}^T \textbf{Z} u \simeq \frac{1}{p + \lambda} n h^2 = \frac{n}{p} ( h^2 )^2
\end{align*}

\bigskip

The $n>p$ scenario :
\begin{align*}
    \textbf{Z}^T \textbf{Z} \simeq n \textbf{I}_p &\Rightarrow \textbf{K}_{\lambda} \simeq \frac{1}{n + \lambda} \textbf{Z}^T \Rightarrow \textbf{K}_{\lambda}^T \textbf{K}_{\lambda} \simeq \left( \frac{1}{n + \lambda} \right)^2 \textbf{Z} \textbf{Z}^T
\end{align*}

\begin{align*}
    &\text{tr}( \textbf{K}_{\lambda}^T \textbf{K}_{\lambda} \times \sigma^2 \textbf{I}_n ) \simeq  ( 1 - h^2 ) \left( \frac{1}{n + \lambda} \right)^2 \text{tr} ( \textbf{Z} \textbf{Z}^T ) \simeq ( 1 - h^2 ) \frac{n}{(n + \lambda)^2} p = ( 1 - h^2 ) \left( \frac{n}{n + \lambda} \right)^2 \frac{p}{n} \\
    &( \textbf{Z} u )^T \textbf{K}_{\lambda}^T \textbf{K}_{\lambda} ( \textbf{Z} u ) \simeq \frac{1}{(n + \lambda)^2} u^T \textbf{Z}^T \textbf{Z} \textbf{Z}^T \textbf{Z} u \simeq \left( \frac{n}{n + \lambda} \right)^2 u^T u \simeq \left( \frac{n}{n + \lambda} \right)^2 h^2 \\
    &u^T \textbf{K}_{\lambda} \textbf{Z} u \simeq \frac{n}{n+\lambda} h^2
\end{align*}

\bigskip

Concatenating those expressions, we eventually get:
\begin{align}
    \text{corr}(  \hat{y}_{te}, y_{te}) \simeq \left\{ \begin{array}{l}
         \frac{ \frac{n}{p} (h^2)^2 }{ \sqrt{ \frac{n}{p} (h^2)^2} \sqrt{1} } = \sqrt{ \frac{n}{p} } h^2 \text{ if } n < p \\
          \frac{ \frac{n}{n + \lambda} h^2 }{ \sqrt{\left( \frac{n}{n + \lambda} \right)^2 \left( \frac{p}{n} ( 1 - h^2 ) + h^2 \right)} \sqrt{1}} = \frac{h^2}{ \sqrt{\frac{p}{n} ( 1 - h^2 ) + h^2} }, \ \text{ otherwise.}
    \end{array} \right.
\end{align}
\end{document}